\documentclass[aps,preprint,superscriptaddress, showpacs, showkeys]{revtex4}\usepackage{graphicx}

\newcommand{\be}{\begin{equation}}
\newcommand{\ee}{\end{equation}}
\newcommand{\bea}{\begin{eqnarray}}
\newcommand{\eea}{\end{eqnarray}}

\newcommand{\rv}{\rho_{vac}}
\newcommand{\pvc}{p_{vac}}
\newcommand{\rt}{\rho_{tot}}
\newcommand{\pt}{p_{tot}}
\newcommand{\rd}{\rho_{dif}}
\newcommand{\rinf}{\rho_{\infty}}
\newcommand{\rz}{\rho_{0}}
\newcommand{\rc}{\rho_{c}}

\newcommand{\rn}{\rho_{n}}
\newcommand{\rnst}{\rho_{n}^{*}}
\newcommand{\rvst}{\rv^{*}}
\newcommand{\pn}{p_{n}}
\newcommand{\wn}{w_{n}}
\newcommand{\Fn}{F_{n}}
\newcommand{\sn}{s_{n}}
\newcommand{\hn}{h_{n}}
\newcommand{\then}{\theta_{n}}

\newcommand{\rst}{\rho_{*}}
\newcommand{\Vst}{V_{*}}

\newcommand{\km}{\mu^{-}}
\newcommand{\kp}{\mu^{+}}
\newcommand{\kpm}{\mu^{\pm}}

\newcommand{\vph}{\varphi}

\newcommand{\bbi}{{\bf i}}

\newcommand{\br}{{\bf r}}

\newcommand{\bz}{{\bf z}}

\newcommand{\bbe}{{\bf e}}

\newcommand{\ai}{_{\alpha}}
\newcommand{\bi}{_{\beta}}
\newcommand{\abi}{_{\alpha\beta}}

\begin{document}

\title{\vspace{0cm}\large\bf
Gravity and Cosmology with Interacting Dark Energy}

\author{A.~S.~Silbergleit}
\email{alex.gleit@gmail.com}
\affiliation{ HEPL, Stanford University, Stanford, CA 94305-4085, USA}

\date{\today}

\begin{abstract}
Dark energy (DE) is not necessarily uniform when other sources of gravity are present: interaction with matter leads to its variation in space and time. We study cosmological implications of this fact by analyzing cosmological models in which DE density interacts with matter and thus changes with time. We model the DE--matter interaction by specifying the rate of change of the DE density as an arbitrary function of it and the density of matter, in a single--phase case. In the case of several matter components interacting with dark energy we assume the rate of every interacting phase density to be an arbitrary function of this density and the DE density. We describe some properties of cosmological solutions valid for a general law of DE--matter interaction, and discuss physical admissibility of the interaction laws. We study numerous families of exact solutions, both singular, non-singular, and mixed. Some of them exhibit interesting properties, such as, for instance, absence of the horizon problem due to the initial fast growth of the scale factor (any power of time possible); non-singular  evolution from one de Sitter universe (pure DE with no matter) to the other one with a different DE density; DE dominating either from some moment of time on, or throughout the expansion; dark matter dominating normal matter at large times without any parameter tuning, and so on. All the results are obtained strictly within the framework of general relativity, Einstein's theory of gravity, without modifying it in any way.
\end{abstract}

\pacs{04}

\maketitle
\vfill\eject
\tableofcontents
\vfill\eject

\section{Introduction. Non-Uniform Dark Energy\label{s1}}

According to Einstein's equations of general relativity, any energy-momentum tensor $T\abi$ must satisfy the condition of energy and momentum conservation,
\be\label{cons}
T^{\alpha}_{\beta;\alpha}=0\; .
\ee
In 1965 Gliner~\cite{Glin65} pointed out that the simplest energy--momentum tensor is that of what is now called the dark energy, or heavy vacuum: 
\be\label{TDE}
T\abi=\rho_{vac}\, g\abi\; .
\ee
Here $\rv$ is the DE density proportional to the Einstein cosmological constant, $\Lambda$.

If dark energy is the {\it only} source of gravity, i.e., there are no other terms in the energy--momentum tensor except the one in the equation~(\ref{TDE}), the condition~(\ref{cons}), in view of 
$g^{\alpha}_{\beta;\alpha}=\delta^{\alpha}_{\beta;\alpha}=0$,  implies
\be\label{rVconst}
\rho_{vac,\,\alpha} =0,\qquad \rho_{vac}=\mbox{const}\; .
\ee
In other words, the density of heavy vacuum is uniform always and everywhere if the spacetime is created only by DE, with nothing else present in it.

Gliner also established the equation of state (EOS) of DE:
\be\label{EOSV}
p_{vac}= - \rho_{vac}\;  ;
\ee
thus for positive DE density its pressure is negative. This EOS is easily verified by comparing the tensor~(\ref{TDE}) with the energy-momentum tensor of a perfect fluid:
\be\label{Tmat}
T\abi=-p g\abi+(\rho+p)u\ai u\bi\; .
\ee
The actual values of density, $\rho$, and pressure, $p$,  are restricted by the equations following form the conservation condition (\ref{cons}) for each particular metric. 

Writing the energy-momentum tensor when DE and other gravity sources are present in the most general form as
\[
T\abi=\rho_{vac}\, g\abi +Q\abi,\qquad Q\abi\not=0\; ,
\]
we stress that the conservation law requires only
\[
\rho_{vac,\,\beta}+Q^{\alpha}_{\beta;\alpha} =0\; .
\]
So $\rho_{vac} \not=\mbox{const}$, unless $Q\abi$ is conserved separately.

The tensor~(\ref{Tmat})  is used in cosmology to describe matter in the universe. Recent observational data from Supernovae and CMB anisotropy demonstrated the dominating presence of DE in our universe. Therefore its complete energy-momentum tensor is now taken as a sum of the tensors~(\ref{TDE}) and~(\ref{Tmat}):
\be\label{Tfull}
T\abi=\rho_{vac}\, g\abi+\left[-p g\abi+(\rho+p)u\ai u\bi\right]\; .
\ee
(in particular, with $p=0$ this is assumed in the $\Lambda$CDM model).
By condition~(\ref{cons}), the divergence of this expression must vanish:
\be\label{consTfull}
\rho_{vac,\,\beta}+\left[-p g^\alpha_\beta+(\rho+p)u^\alpha u_\beta\right]_{;\,\alpha}
=0\; .
\ee
Nothing else is implied by the conservation condition~(\ref{cons}): as soon as~(\ref{consTfull}) is true, energy and momentum are conserved, and vice versa. 

However, for no evident reason, except simplicity, computational convenience, and perhaps some kind of intellectual inertia, in modern cosmology it is usually assumed that DE density is uniform, as in~(\ref{rVconst}). So each of the two terms describing DE and matter is assumed to be conserved {\it separately}, and matter density and pressure satisfy the same conservation  equations as in the case when there is nothing else but matter.

Apparently, this assumption is mathematically redundant; even worse, it is suspicious from the physics standpoint, because it makes heavy vacuum absolute, independent of anything else, by forbidding, in fact, its interaction with matter. We do not see why should it be so; rather, it seems natural to think that dark energy, being a special state of a physical medium, should interact with other physical substances populating the universe. Gliner~\cite{Glin69} and Gliner and Dymnikova ~\cite{GlD75} held this point of view, but did not  pursue in full its cosmological implications. Their most important suggestion of a non-singular cosmology could not reflect the modern knowledge of the strong continuous presence of heavy vacuum, thus considering DE as just a non-singular initial state of the universe that turns to pure matter through an instant phase transition at the beginning of the cosmological expansion.

Along these lines of thinking, below we study in detail the Friedmann cosmology with variable density of heavy vacuum, that is, under the condition~(\ref{consTfull}) only. In it, DE and matter coexist and permanently interact with each other. The interaction is modeled in a rather general way strictly within the framework of general relativity.

This appears to be even more reasonable since the $\Lambda$CDM model  is found in certain contradictions (`tensions') with the modern observational data (see paper~\cite{Buch&Co} and the references therein). Moreover, recent observations tentatively indicate that dark energy in our universe does evolve (see e.g.~\cite{SSStar}), so some alternatives to the constant DE model have been considered in papers~\cite{Overd98}~-~\cite{FDS} (see also the references therein and in~\cite{SSStar}).

\section{Friedmann Cosmology with Changing Dark Energy\label{s2}}

The energy-momentum tensor~(\ref{Tfull}) can be written in the form~(\ref{Tmat}) of a perfect fluid,
\be\label{TLamCDM}
T\abi=-\pt\, g\abi+(\rt+\pt)u\ai u\bi\; ,
\ee
whose density, $\rt$, and pressure, $\pt$, are defined as
\be\label{rpt}
\rt=\rho+\rv,\qquad \pt=p+\pvc=p-\rv\; ;
\ee
the last expression here is implied by the DE equation of state~(\ref{EOSV}).

We study the Friedmann cosmology using the Robertson -- Walker metric
\be\label{FRW}
ds^2=dt^2-a^2(t)\left[\left(1-kr^2\right)^{-1}dr^2+d\theta^2+r^2\sin^2\theta \,d\vph^2\right]\; ,
\ee
with the dimensionless scale factor $a(t)$ being the only unkown, and $k=0,\,-1,\,1$ for the flat, open and closed universe, respectively; we use the system of units with $c=G=1$. 

For the expressions~(\ref{TLamCDM}) and (\ref{FRW})  Einstein's equations are known to reduce to the Friedmann equations, which we write as:
\be\label{FrEq}
3\left(\frac{\dot a}{a}\right)^2=8\pi\rt-\frac{k}{a^2};\qquad \dot\rt=-(\rt+\pt)\,3\,\frac{\dot a}{a}\; 
\ee
(the dot always denotes the derivative in time). The second of this equations is one of the four conditions~(\ref{consTfull}) in co-moving coordinates ($u_0=1,\;u_1=u_2=u_3=0$) with $\beta=0$; it expresses  energy conservation in a co-moving volume (see below). The other three conditions~(\ref{consTfull}) require that $\rv$ is independent of all spatial coordinates, which is also clear from the assumptions made.

Equations~(\ref{FrEq}) can be combined to yield the expression for the acceleration,
\be\label{acc}
3\,\frac{\ddot {a}}{a}=-4\pi\,(\rt+3\pt)\; .
\ee
It shows that the expansion accelerates, decelerates, or proceeds uniformly depending on the sign of the `effective gravitating density' $\rt+3\pt$ (negative, positive or zero, respectively). If  only DE is present, i.e., $\rt=\rv,\;\pt=\pvc=-\rv$, then
\[
\rt+3\pt=-2\rv<0\; ,
\]
and expansion accelerates; thus heavy vacuum gravity is repulsive. In the opposite case of pure matter, $\rv=0$, the sign depends on its equation of state; usually matter is attractive, leading to deceleration.

It is convenient to introduce the co-moving volume as
\be\label{covol}
V(t)=a^3(t),\qquad a(t)=V^{1/3}(t)\; ,
\ee
and rewrite the Friedmann equations~(\ref{FrEq}) in terms of it. Using $V(t)$ in the second equation as an independent variable  instead of time, we obtain:
\be\label{FrEqV}
\frac{1}{3}\,\left(\frac{\dot V}{V}\right)^2=8\pi\rt-\frac{k}{V^{2/3}};\qquad\
\frac{d\left(\rt V\right)}{dV}= -\,\pt\; .
\ee

The first of these equations determines the time dependence of the volume, and hence of the scale factor and all other parameters. Indeed, as soon as the total density is known as a function of the volume, $\rt=\rt(V)$, the dependence of the latter on time is obtained by direct integration, namely:
\be\label{volt}
t-t_0=\int\limits_{V_0}^{V(t)}\,\frac{dV}{V\sqrt{3\left[8\pi\rt(V)-k\,V^{-2/3}\right]}},\qquad V_0=V(t_0)\; .
\ee

Thus everything reduces to the second of equations~(\ref{FrEqV}), which describes conservation of total energy, $\rt V$,  in the co-moving volume by implying
\[
d\left(\rt V\right)+\pt\,dV=0\; .
\]
This single equation, however, contains three unknown functions, $\rho,\;\rt$ and $p$:
\be\label{keyEq}
\frac{d\left[(\rv+\rho)V\right]}{dV}= -\,\pt=-(p+\pvc)= -(p-\rv)\; .
\ee
The equation of state (EOS) of a single--phase matter relates its pressure and density, reducing thus the number of unknowns to two, $\rho$ and $\rv$  Of course, it is impossible to determine both of them simultaneously form the single equation~(\ref{keyEq}). The situation is even worse when there are $N>1$ matter components, each with its own EOS; then all the $N+1$ densities are unknown, with just the same single equation for all of them (see sec.~\ref{s4}).

Clearly, what is missing yet is the law of interaction between DE and matter, which would provide the second equation needed to determine the expansion completely. Its physical derivation, especially from the first principles, is an outstanding problem of physics and cosmology, and a great challenge to the physics theory. Since we currently do not know how to derive this equation, the only way to understand possible features of the universe seems to rely on certain plausible models (and hope that at least some of them are not very far from reality!).

In what follows we model the interaction of heavy vacuum with matter, and study cosmological solutions that stem form these models; some of the solutions exhibit remarkable properties. We will use the energy conservation equation~(\ref{keyEq}) in the form more convenient for our purposes:
\be\label{key1}
\frac{d\left[(\rv+\rho )\right]}{dV}=  -\frac{(\rho+p)}{V}\; .
\ee

\section{Cosmology with  Dark Energy and a Single Type \\ 
of Matter: General Interaction Model \label{s3}}

Let matter be present  in a single phase with the equation of state  
\be\label{EOS1ph}
p=w\rho \; .
\ee
Since $\rho+3p= (1+3w)\rho$, the matter is attractive, according to~(\ref{acc}), when\break  $w>-1/3$ \footnote{Matter with the EOS where $w$ is negative, but larger than $-1$, is usually called quintessense. Its gravity is attractive for $-1/3<w<0$, repulsive for $-1<w<-1/3$, and `neutral' when $w=-1/3$ (no gravitational acceleration). Since Einstein effectively used neutral quintessense in his static cosmological model of 1917, we named it ~`Einstein's quintessense' in our paper~\cite{CSS}.}. However, as a rule this parameter is non-negative for the known types of matter: $w=0$ for pressure-less matter (`dust'), $w=1/3$ for radiation (ultra-relativistic gas), and $w=1$ for the super-dense Zel'dovich fluid~\cite{Zeld} (it seems the largest $w$ known so far). Eliminating $p$ from equation~(\ref{key1}) by the EOS~(\ref{EOS1ph}), we write it as
\be\label{key11ph}
\frac{d\left[(\rv+\rho )\right]}{dV}=  -\frac{(1+w)\rho}{V}\; .
\ee

Note that if there is no matter, $\rho=0$, then this equation gives $\rv=\mbox{const}$. which is the de Sitter solution. If, on the contrary, only matter is present, $\rv=0$, equations~(\ref{key11ph}) and~(\ref{volt}) imply the usual solutions (see below). When both the DE and matter are present,  the standard approach is to assume that matter conserves separately, $d\rho/dV=-(1+w)\rho/V$. Equation~(\ref{key11ph}) then implies the constant DE density, and the whole solution becomes
\be\label{standsol1}
\rv=\mbox{const},\qquad
\rho=C/V^{(1+w)},\qquad C>0\; .
\ee

We assume, instead, that matter and heavy vacuum are interacting. We model this interaction by specifying the rate of change of the DE density as
\be\label{key21ph}
\frac{d\rv}{dV}= \frac{ F(\rv,\rho)}{V}\; ,
\ee
Here $F$ is some function of two variables, so far arbitrary. Note that it must actually depend on the second variable, $\rho$, otherwise there will be no DE--matter interaction, just an independent law of the evolution of the dark energy density.

Some cosmological models with dynamical dark energy have been studied earlier (see~\cite{Overd98}, ~\cite{Overd99} the references there in and below. A full list of models considered before 1998 is given by  in the paper~\cite{Overd98} see Table I there). Mostly, these models treated the DE density as a known function of time, or the scale factor, or  the Hubble parameter, $H=\dot{a}/a$ \footnote{The term Hubble {\it parameter} is, in fact misleading, since it is a fucntion of time, $H=H(t)$. So it would be better to call it the Hubble {\it function}; in this paper, however, we go along with the universal usage.}, or the acceleration parameter $\ddot{a}/a$ (typically, some power functions have been used). The first two dependencies do not actually model the DE--matter interaction, rather try to trace it consequences.

A cosmological model for a flat Friedmann universe with a single matter phase and dynamical dark energy whose density is a known function of the Hubble parameter, $\rv=\rv(H),\;H=\dot{a}/a$, was systematically studied by I.L. Shapiro and J.~Sol\`a and their co-authors in papers~\cite{ShapSol}~-~\cite{LimBasSol}. They took into account some considerations of the renormalization group techniques of quantum field theory hinting that  $\rv(H)$ might be a series in even powers of $H$; particular solutions were studied  with DE density being an even polynomial of $H$. 

Our model~(\ref{key21ph}) allows for an arbitrary $\rv(H)$ as its particular case.  Indeed, specifying the interaction function as 
\[
F(\rv,\rho)=F(\rv+\rho)=F(\rt)\;,
\]
 by the first of the Friedmann equations~(\ref{FrEq}) with $k=0$ we find that it is a function of the Hubble parameter only, $F=\Phi(H)$. Function $\Phi(H)$ is determined from the compatibility of equations~(\ref{FrEq}) and~(\ref{key21ph})  as soon as $\rv(H)$ is fixed; the details are given in Appendix~\ref{O}. Still, this model, in which the DE--matter interaction is completely determined by the total density only, appears to be not very compelling. However,  this is not quite  so for the open and closed universe.  In  Appendix~\ref{O} we extend the $\rv(H)$ model to these cases, $k=\mp1$, and show that it corresponds to some complicated enough interaction function $F(\rv,\rho)$.
 
Finally, the model with $\rv=\rv(\ddot{a}/a)$ is also a particular case of our model~(\ref{key21ph}), in view of the acceleration equation~(\ref{acc}). In this case the interaction function is
\[
F(\rv,\rho)=\Phi(\eta),\quad \eta=\rt+3\pt=(1+3w)\rho-2\rv=(-3/4\pi)(\ddot{a}/a)\; ;
\]
it is related to $\rv(\eta)$ by the equality
\[
\Phi(\eta)=-\frac{(1+w)[\eta+2\rv(\eta)]\rv^{'}(\eta)}{1+3(1+w)\rv^{'}(\eta)}\; .
\]
This is similar to the expressions~(\ref{O6}) and~(\ref{O11}) of the previous model $\rv(H)$, and it works for a cosmology with any spacetime curvature, $k=0,\pm1$.

 Returning to our consideration we notice that as soon as $F(\rv,\rho)$ is specified, the system of two equations~(\ref{key11ph}) and~(\ref{key21ph}) allows one to determine $\rv$ and $\rho$ as functions of $V$, and, due to~(\ref{volt}), as function of time, i.e., to get the complete picture of cosmological expansion. The form~(\ref{key21ph}) of the interaction equation is rather general; on the other hand, it simplifies the choice of a particular model belonging to this wide class. It is straightforward to combine the equations so that each of them contains just one  derivative (standard form): 
\be\label{govs1ph}
\frac{d\rho}{dV}=  -\frac{(1+w)\rho+ F(\rv,\rho)}{V},\qquad\qquad
\frac{d\rv}{dV}= \frac{ F(\rv,\rho)}{V}\; .
\ee
Introducing a new independent variable $\lambda=\ln\left(V/V_*\right)$ ($V_*=\mbox{const}>0$ is arbitrary), we see that  the  governing system is {\it autonomous}:
\be\label{govsaut1ph}
\frac{d\rho}{d\lambda}=  -\left[(1+w)\rho+ F(\rv,\rho)\right];\quad
\frac{d\rv}{d\lambda}= F(\rv,\rho),\qquad  \lambda=\ln\left(V/V_*\right)\; .
\ee

Thus a powerful arsenal of methods applicable to autonomous systems in the plane can be used for a qualitative study of solutions of equations~(\ref{govsaut1ph}) based on the properties of the interaction function $F(\rv,\rho)$. Alternatively, we will investigate particular implementations of the general model and study the properties of the corresponding exact solutions.

We are only interested in physically meaningful solutions for which the densities $\rho,\;\rv$ are non-negative, and the matter density vanishes at large times:
\be\label{rhoto0}
\rho\to+0\quad\mbox{when}\quad \lambda\to+\infty\quad(V,\;a,\;t\to+\infty)\;. 
\ee
It is also natural to assume that the DE density is bounded at large times for a physically sound cosmological solution. Thus we exclude "the big crunch" from the discussion, mainly to limit our rather extended analysis. Also, a typical solution has either an initial, or a final singularity, or none at all, but not both simultaneously; the existence of the big crunch in our universe seems rather improbable.

The above requirements restrict the possible interaction laws. For instance, no physical solutions exist when the interaction function is bounded away from zero for all relevant values of its arguments,
\[
F(\rv,\rho)\leq - F_0<0, \quad\mbox{or}\quad F(\rv,\rho)\geq  F_0>0;\qquad 
F_0=\mbox{const}>0\; .
\]
Indeed, in the first case the DE density becomes negative at a finite moment of time and tends to the negative infinity in the large time limit; in the second case it goes to the positive infinity; and in both cases the matter density does not tend to zero at large times. This is easily seen, under the above conditions, from the equations~(\ref{govsaut1ph}).

The way to meet the requirement~(\ref{rhoto0}) is to have an attracting rest point \break $\rho=0,\;\rv=\rinf=\mbox{const}>0$. All the rest points (critical points,  equilibria) of the system~(\ref{govsaut1ph}) are described by the equations
\be\label{DeSit}
\rho=0,\qquad F(\rv,0)=0\; .
\ee
A physical  rest point exists when the second of these equations has a non-negative root. If the root is zero, then no source of gravity is present, and it is a Minkowsky spacetime; a positive root corresponds to a de Sitter universe with the uniform DE and no matter. (Quite appropriately, the de Sitter universe is static, even though it can be described by the Robertson---Walker metric with the time--dependent scale factor). So, if a cosmological solution tends to such rest point, then the final state is a de Sitter universe, as in the $\Lambda$CDM model.
 
We now go about some particular interaction models and analyze the exact cosmological solutions emerging from them.

\subsection{Linear  Interaction between Dark Energy and Matter. \\ Exact Solution for Singular Cosmology Dominated by Dark Energy\label{s4}}

The simplest kind of DE--matter interaction appears to be when the rate of $\rv$ is proportional to $\rho$,
\be\label{Mod1F}
 F(\rv,\rho)=-s\rho\;.
\ee
This is our first choice; it is remarkable also because it introduces just one new dimensionless parameter,  $s$. In addition, the whole positive semi--axis $\rv\geq0$ consists of the roots of the equation~(\ref{DeSit}), so that $\rho=0,\;\rv=\rst\geq0$ is  the rest point for any $\rst$.

The governing equations~(\ref{govsaut1ph}) become
\[
\frac{d\rho}{d\lambda}=  -(1+w-s)\rho;\qquad
\frac{d\rv}{d\lambda}= -s\rho\; ,
\]
The first of them shows that this type of interaction with DE just changes the parameter in the linear EOS~(\ref{EOS1ph}) of matter, replacing $w$ with $w-s$. In other words, the effective pressure here is
\be\label{peff}
p_{eff}= (w-s)\,\rho\; ;
\ee
noteworthy, matter effectively acts as quintessense when $s>w$, and its gravity  becomes repulsive for $s>w+1/3$. 

The governing linear system with constant coefficients is  immediately integrated; the result for $s\not=1+w$, in terms of the independent variable $V$, is:
\bea\label{Mod1sol}
\rho=\frac{C}{V^{1+w-s}}\; ;\qquad\qquad\qquad\qquad\nonumber\\
\rv=\rho_\infty +\frac{s}{1+w-s}\,\frac{C}{V^{1+w-s}}=\rho_\infty +\frac{s}{1+w-s}\,\rho\; ;\\
\rt=\rho_\infty +\frac{1+w}{1+w-s}\,\frac{C}{V^{1+w-s}}=\rho_\infty +\frac{1+w}{1+w-s}\,\rho\; ,\nonumber
\eea
where $C>0,\;\rinf\geq0$ are arbitrary constants. The evolution of matter density is given by  a power dependence, but the power value is different than the usual one because of the DE--matter interaction.

Condition~(\ref{rhoto0}) of vanishing $\rho$ at large times restricts parameter $s$ from above: $s<1+w$ (recall that we excluded the big crunch from our discussion). The lower bound for it comes from the requirement $\rv>0$ implying $s>0$ (otherwise the DE density is negative at least for small enough times, $V\to+0$). Therefore the interaction parameter $s$ proves to be well constrained:
\be\label{srange}
0<s<1+w \; .
\ee

The left inequality here means, by the way, that the interaction is permanently reducing dark energy and producing matter; there is no physical solution in the opposite case. Notably, here not only the matter, but also the DE density is singular at the beginning. So both matter and dark energy are born in the Big Bang, unlike the usual solution with uniform dark energy, where the density of the latter is some finite constant.
 
The time behavior of the solution is given by the formula~(\ref{volt}) where it is convenient to set $t_0=0$ and
 $V_0=0$. The integral there can be calculated explicitly in some cases (see~\cite{AS}, where similar integrals are treated systematically). Here we show just most significant small and large time asymptotics of the solution~(\ref{Mod1sol}), which is straightforward to get from the analysis of the expression~(\ref{volt}). Evidently, for large time ($V\to+\infty$) we have the usual exponential acceleration caused by the limit constant DE density, $\rho_\infty$:
 \bea\label{Mod1tlarg}
 a\sim\exp(t/\tau),\quad\rho=O(\exp(-3(1+w-s)t/\tau))\to0\;, \nonumber\\
 \rv\to\rho_\infty+O(\exp(-3(1+w-s)t/\tau))\to\rho_\infty,\quad
  t\to+\infty\nonumber\; ;\\
  \tau= \sqrt{8\pi\rho_\infty/3}.\qquad\qquad\qquad\qquad\qquad
 \eea
 
 The solution emerges from the initial singularity ($V\to+0$) according to
 \be\label{Mod1tsmal}
 a\sim t^{2/3(1+w-s)}\to+0,\quad\rho\sim\rv\sim t^{-2}\to+\infty,\quad t\to+0\;. \nonumber\\
 \ee
Of course, for $s=0$ the scale factor and matter density behave exactly as in the Friedmann solutions, since there is no DE--matter interaction. For the values of the interaction parameter $s$ from the physical range~(\ref{srange}), the power $2/3(1+w-s)$ that specifies the initial time dependence of the scale factor is always larger than its Friedmann value $2/3(1+w)$. The expansion thus goes faster, at least in the beginning. This is the effect of the repulsive heavy vacuum  whose density is as singular as the matter density is. Moreover, this power can be larger than unity, $2/3(1+w-s)>1$ when $s>w+1/3$, so that for
\be\label{Mod1horiz}
w+1/3<s< w+1\; 
\ee 
 there is no horizon problem. Consistently, this is the range where the existing material effectively behaves as a repulsive quintessense, since, by (\ref{peff}), its effective gravitating density becomes negative,
\[
\rho+3p_{eff}=3(w+1/3-s)\rho<0\; .
\] 
The expansion is faster, the closer $s$ is to $w+1$, it beats any power of time when $s$ tends to this upper bound of its range. One can speak thus about `inflation', but of the power, rather than the exponential, one.

Note that the parameter range~(\ref{Mod1horiz}) is impossible for the closed universe\break  ($k=1$) requiring $0<s\leq w+1/3$, to compensate for the negative curvature term\break  $-a^{-2}=-V^{-2/3}$ (see equation~(\ref{volt})).
 
Quite naturally, Friedmann solutions ($\rv=0$) cannot be obtained from~(\ref{Mod1sol}), because  for $\rv=0, \rho\not=0$ the second of the governing equations is contradictory unless $\rho\equiv0$. On the other hand, the mentinoned de Sitter solution with any value of the DE density is given by the expressions~(\ref{Mod1sol}) with $C=0$.
 
 Next we note that the solution~(\ref{Mod1sol}) is dominated by DE at large times independent of parameter values. To see what is dominating for other periods of evolution, it is instrumental to calculate the difference
  \be\label{Mod1dsif}
\rd=\rv-\rho=\rho_\infty +\frac{2s-(1+w)}{1+w-s}\,\frac{C}{V^{1+w-s}}=
\rho_\infty +\frac{2s-(1+w)}{1+w-s}\,\rho\; .
 \ee
 This formula shows that the range~(\ref{srange}) of the interaction parameter $s$ is divided by exactly its midpoint, $(w+1)/2$, into two parts:  $0<s<(w+1)/2$, with the corresponding cosmological solutions initially dominated by matter, and \break $(w+1)/2\leq  s<w+1$, when DE dominates throughout the expansion. Indeed, in the latter case $\rd>0,\;\rv>\rho$ at all times (and $\rd$ even turns to $+\infty$ when $t\to+0$ for $s$ strictly larger than $(w+1)/2$). In the former case the density difference is negative ($\rho>\rv$, matter dominates) from the beginning until
\[ 
 V=\left[\frac{1+w-2s}{1+w-s}\,\frac{C}{\rho_\infty}\right]^{\frac{1}{1+w-s}},\qquad 0<s<(w+1)/2\; ,
 \]
when it turns to zero. After this moment the difference becomes positive, DE starts dominating and continues for the rest of the time.
 
 Summarizing our last observations we point out that: a) the initial expansion can be fast enough to resolve the horizon paradox for $s>w+1/3$; b) DE is dominating throughout the whole expansion if and only if
 \be\label{Mod1fastDEd}
s\geq (w+1)/2\; 
\ee
(this means $s\geq 1/2$ for dust ($w=0$), and $s>2/3$ for radiation ($w=1/3$)).

We also note that the excluded case $s=1+w$ produces a physically meaningless but rather peculiar solution:  the matter density does not change, $\rho=\mbox{const}$, while the DE density goes from  plus to minus infinity as a log of the inverse scale factor.

An interaction depending on an arbitrary linear combination of both densities is analyzed below in section~\ref{s6.1}.

\subsection{The Role of Non-Linearity: Interaction Laws $F(\rv,\rho)=f(\rho)$ \\ and the Corresponding  Class of Exact  Solutions\label{s5}}

\subsubsection{General Solution and Its Properties\label{s5.1}}

A natural generalization of the linear interaction law~(\ref{Mod1F}) is:
\be\label{GenF(rho)}
 F(\rv,\rho)=f(\rho)\; ,
\ee
where $f$ is an arbitrary function of one argument. The first of the governing equations~(\ref{govs1ph}) again is the equation for matter density only, which shows that the interaction~(\ref{GenF(rho)})  means the change in the matter EOS from the linear relation ~(\ref{EOS1ph}) to the following  non-linear one:
\be\label{EOSnl}
p_{eff}=w\rho+f(\rho)\; .
\ee

Integration of the separable equation~(\ref{govs1ph}) provides an algebraic equation for the matter density $\rho=\rho(V)$, 
\be\label{Genrho}
\ln\frac{V}{V_*}=-\int\limits^\rho\,\frac{dx}{(1+w)x+f(x)}\; .
 \ee 
Then the DE density is found by integrating a known function. This  operation, however, might be not easy to practically implement when the equation~(\ref{Genrho}) cannot be solved for $\rho$ explicitly.  To overcome this difficulty, we note that the governing equations imply
 \[ 
 \frac{d\rv}{d\rho}=-\frac{f(\rho)}{(1+w)\rho+f(\rho)}\; ,
 \]
 giving an expression for $\rv$ through $\rho$, $\rv=\rv(\rho(V))$:
 \be\label{Genrhovac}
\rv=-\int\limits^\rv\,\frac{f(x)\,dx}{(1+w)x+f(x)}\; .
 \ee 

Depending on the properties of the function $f$, formulas~(\ref{Genrho}) and~(\ref{Genrhovac}) might or might not represent {\it physical} solutions that require  both densities to be positive and have a reasonable behavior. Even the condition~(\ref{rhoto0}) of matter density vanishing at large times might not be fulfilled; however, if $f(\rho)$ behaves linearly for small $\rho$,
\be\label{frosmall}
f(\rho)=-s\rho\,[1+o(1)], \qquad \rho\to+0,
\ee
then the expression~(\ref{Mod1sol}) for matter density is retained for large times (large $V$), so  it turns to zero at the end of expansion.
 
To specify $f(\rho)$ in such way that the obtained solution is physically meaningful turns out to be not an easy job. For instance, let us take a non-linear interaction described by a quadratic dependence,
 \be\label{Mod2F}
 F(\rv,\rho)=f(\rho)=-\rho^2/R\; ,
\ee
where  $R$ is an arbitrary parameter. Calculating the integral~(\ref{Genrho})  results in the following expression for the matter density:
\bea\label{Mod2sol}
\rho=\frac{(1+w)R}{(V/\Vst)^{1+w}-1}\,\left(\frac{V}{\Vst}\right)^{1+w},\quad\rho>(1+w)R\; ;\nonumber\\
\rho=\frac{(1+w)R}{(V/\Vst)^{1+w}+1}\,\left(\frac{V}{\Vst}\right)^{1+w},\quad\rho<(1+w)R\; .
\eea
It might look nice, but is not relevant, in fact, from the physics point of view, for both signs of parameter $R$. It is straightforward to see that a meaningful  positive solution does not exist on the whole semi-axis $0<V<\infty$, and it has a strange singularity at  {\it a finite time} (finite volume $V=\Vst$). Moreover, if the physical solution exists at large times, it does not go to zero when $t\to+\infty$, tending instead to the positive limit $(1+w)R$.

As it turns out, same problems surface for any non-negative power laws $f(\rho)\propto\rho^\mu,\;\mu>1$, as well as for the inverse power dependencies $f(\rho)\propto\rho^{-\mu},\;\mu>0$. Polynomial functions, like
\[
 f(\rho)=-s\rho-\rho^2/R\; ,
\]
satisfying condition~(\ref{frosmall}) do not help, either, leading to non-physical solutions. 

All these laws do have something in common:  for all of them the non-linearity becomes overwhelming, too strong, in one of the characteristic limits $\rho\to+0$ or $\rho\to+\infty$. This observation brings one to a thought that condition~(\ref{frosmall})  could be helpful in {\it both} limits of small and large densities,
\be\label{frosmallarg}
f(\rho)=-s\rho\,[1+o(1)], \qquad \rho\to+0\quad\mbox{and }\quad \rho\to+\infty\;.
\ee
And indeed, asymptotic analysis of the integrals~(\ref{Genrho}) and~(\ref{Genrhovac}) under this condition immediately demonstrates that the solution~(\ref{Mod1sol}) is retained, for small and large $\rho$ (small and large $V$), and thus the proper behavior~(\ref{Mod1tlarg}) and~(\ref{Mod1tsmal}) takes place.

Condition~(\ref{frosmallarg}) subordinates non-linear part of interaction to the linear one for both small and large densities. However, between the start and end of the expansion, non-linearity can be dominating and cause thus significant deviations of parameters from their values~(\ref{Mod1sol}) obtained for purely linear interaction. Moreover, the strong influence of non-linearity can lead to new singularities in the course of the expansion, again making the corresponding solution non-physical. For this reason, ~(\ref{frosmallarg}) is {\it not a sufficient} condition for a solution to be sound from a physical standpoint; it only guarantees an acceptable behavior at initial singularity and infinity.

We illustrate these peculiarities using one but rich enough example. 

\subsubsection{ Example: Exact Solution for a Special Form of Function $f(\rho)$\label{s5.2}}

Let us consider the interaction law~(\ref{GenF(rho)}) with
\be\label{fronlex}
f(\rho)=-s\rho\,\left(1+\frac{\theta\rho^{1/2}}{\rho+R}\right)\; ;
\ee
here $\theta$ is a new parameter. This function satisfies condition~(\ref{frosmallarg}):
\[
f(\rho)=-s\rho\,[1+O(\rho^{1/2})], \;\; \rho\to+0; \quad
f(\rho)=-s\rho\,[1+O(\rho^{-1/2})],\;\; \rho\to+\infty\; .
\]
The  integrands of integrals~(\ref{Genrho}) and~(\ref{Genrhovac}) turn into rational fractions by the substitution  $x=\sqrt{\rho}$, thus both integrals are calculated explicitly in terms of elementary functions. The result depends on the behavior of the following quadratic polynomial in the denominator of some fractions,
 \be\label{pquad}
 P(x)=x^2-qx+R,\qquad q=s\theta/(1+w-s)\; ,
 \ee
that is, on the sign of its discriminant 
\be \label{diskr}
\Delta=q^2-4R\; .
\ee

\noindent\underline{Case 1: ${\Delta=q^2-4R<0}$\label{s5.2.1}}  

Since the non-linearity of interaction~(\ref{fronlex}) is stronger, the larger $\theta$ and the smaller $R$, here we deal with a (relatively) weak non-linearity. In this case the polynomial $P(x)$ is positive on the whole real axis, and formulas~(\ref{Genrho}) and~(\ref{Genrhovac}) reduce to the equations ($C>0,\;\rst>0$ are constants of integration):
\bea\label{Exsol1rho}
\rho=\frac{C}{V^{1+w-s}}
\exp{\left[-q|\Delta|^{-1/2}\arctan{\xi(\rho)}\right]},\quad \xi(\rho)=\left(\rho^{1/2}-q/2\right)|\Delta|^{-1/2}\; ;\\ 
\label{Exsol1rv}
\rv=\rst+\frac{s}{1+w-s}\rho+2\theta\,\frac{s(1+w)}{(1+w-s)^2}\left[\rho^{1/2}+h(\rho,s,\theta,R,w)\right]\;,\qquad\\
h(\rho,s,\theta,R,w)=\left(q^2/2 - R\right)|\Delta|^{-1/2}\arctan{\xi(\rho)}+(q/2)\ln{\left[1+\xi^2(\rho)\right]}\nonumber\; .
 \eea 
Equation~(\ref{Exsol1rho}) for the matter density has a unique solution $\rho=\rho(V)$ in the whole range $0<V<\infty$, which behaves properly, i. e., starts with a singularity at $V=0$ and monotonically goes down to zero at infinity. Therefore the DE density~(\ref{Exsol1rv}) also behaves properly.

Both expressions, though somewhat cumbersome, do not differ much qualitatively from their counterparts in the `basic' solution~(\ref{Mod1sol}). Particularly, they require the same parameter range~(\ref{srange}), $0<s<1+w$. The exponential factor in~(\ref{Exsol1rho}) replaces unity in the corresponding equation~(\ref{Mod1sol})  and, since the arc-tangent is bounded, it does not change much throughout the expansion. However, depending on parameters, the difference in the matter density values~(\ref{Exsol1rho}) and~(\ref{Mod1sol}) can be large for some period of time, especially when $q^2$ is close to $4R$. 

 As seen from~(\ref{Exsol1rv}), the DE density also behaves in a familiar way: the two first terms are same as in~(\ref{Mod1sol}), followed by the correction due to interaction non-linearity proportional to $\theta$. Also as in the case~(\ref{Mod1sol}), the solution is permanently dominated by dark energy when $s>(1+w)/2$. The main difference as compared to $\rv$ from the solution~(\ref{Mod1sol}) is: 1) the constant of integration $\rst$ is not the limit value of the DE density at infinity, due to the addition from the non-linearity,
\[
\rinf=\rst+\left(R-q^2/2\right)|\Delta|^{-1/2}\arctan{\left(q|\Delta|^{-1/2}/2\right)}\; ;
\]
2) according to the equation~(\ref{Exsol1rv}), non-linearity adds the log term to the initial singularity in the DE density; it is, however, weaker than the power singularity. 

Anyway, this is a physically meaningful solution; not always so in other cases.
\vfill\eject
\noindent\underline{Case 2: ${\Delta=q^2-4R=0,\;q^2=4R}$\label{s5.2.2}}  

 Here we deal with the boundary case between weak and strong interaction non-linearity. The equations for the densities have the same structure~(\ref{Exsol1rho}),~(\ref{Exsol1rv}):
\bea\label{Exsol2rho}
\rho=\frac{C}{V^{1+w-s}}
\exp{\left(\frac{q}{\rho^{1/2}-q/2}\right)}\; ;\qquad\qquad\qquad\quad\\ 
\label{Exsol2rv}
\rv=\rst+\frac{s}{1+w-s}\rho+2\theta\,\frac{s(1+w)}{(1+w-s)^2}\left[\rho^{1/2}+h(\rho,s,\theta,R,w)\right]\;,\\\
h(\rho,s,\theta,R,w)=q\left[\frac{q^2/4}{\rho^{1/2}-q/2}-\ln\frac{\rho^{1/2}-q/2}{q/2}\right]\nonumber\; .\qquad\qquad
 \eea 
The rest, however, totally depends on the sign of the interaction parameter $q$ coinciding with the sign of $\theta$ for the required range $0<s<1+w$.
 
 If it is negative, $q=-2\sqrt{R}<0$, then equation~(\ref{Exsol2rho}) has a unique solution $\rho(V)$ for any positive $V$, and both densities exhibit exactly the same qualitative behavior as in the case a). All observations made regarding the expressions~(\ref{Exsol1rho}),~(\ref{Exsol1rv}) remain true, we have yet another physically meaningful solution. Its matter density monotonically decreases from infinity at $t=0\;(V=0)$ to zero at $t=\infty\;(V=\infty)$, and positive DE density goes from infinity to some positive value. This happens because the non-linear part of the law~(\ref{fronlex})  works {\it against} the linear one, reducing  thus the DE--matter interaction, i.e., making it weaker. In other words, the effective pressure~(\ref{EOSnl}) corresponding to 
the interaction~(\ref{fronlex}) ,
\[
p_{eff}= (w-s)\rho +\frac{s|\theta|\rho^{1/2}}{\rho+R}
\]
is less repulsive due to non-linearity.

In the opposite case $q=2\sqrt{R}>0$ the picture is drastically different. An essential singularity at $\rho=q^2/4$ appears in the equation~(\ref{Exsol2rho}) which prevents it from having a unique solution with the reasonable behavior on the whole semi-axis (note that the graphic analysis of all the functional equations for $\rho$ we obtained so far is rather transparent). Similar to the case of the interaction law~(\ref{Mod2F}) and other ones mentioned above, there is either a solution on a finite interval $0<V<\Vst$, or a solution on the semi-axis $0<V<\infty$ that starts with zero at $V=0$ and becomes singular at infinity. Thus this case presents no physical solution at all; the interaction proves to be too strong, since its non-linear part enhances the linear one or, in other terms, the corresponding effective pressure
\[
p_{eff}= (w-s)\rho -\frac{s|\theta|\rho^{1/2}}{\rho+R}
\]
turns out too repulsive because of the second term.

\noindent\underline{Case 3: ${\Delta=q^2-4R>0}$\label{s5.2.3}}  

We finally treat the `strongly' non-linear case. The quadratic polynomial~(\ref{pquad}) has two real roots $x_1<x_2$,
\[
x_{1,2}=\frac{1}{2}\left(q\mp\sqrt{\Delta}\right)\; ,
\]
which are both negative for $q<0$, and both positive in the opposite case. The formulas for the densities are
\bea\label{Exsol3rho}
\rho=\frac{C}{V^{1+w-s}}
\left|\frac{\rho^{1/2}-x_2}{\rho^{1/2}-x_1}\right|^{2q/\sqrt{\Delta}}\; ;\qquad\qquad\qquad\qquad\quad\\ 
\label{Exsol3rv}
\rv=\rst+\frac{s}{1+w-s}\rho+2\theta\,\frac{s(1+w)}{(1+w-s)^2}\left[\rho^{1/2}+h(\rho,s,\theta,R,w)\right]\;,\quad\\\
h(\rho,s,\theta,R,w)=
\frac{2}{\sqrt{\Delta}}\left[(R-qx_2)\ln|\rho^{1/2}-x_2|+(R+qx_1)\ln|\rho^{1/2}-x_1|\right]\nonumber\; .
 \eea
They have the same properties as those from the previous case: for $q<0\;(x_{1,2}<0)$ (interaction reduced by non-linearity), equation~(\ref{Exsol3rho}) has no singularities giving rise to a unique meaningful cosmological solution with the kind of behavior described several times above. Contrary to this, when $\theta>0\; (x_{1,2}>0)$  (interaction enhanced by non-linearity), no physical solution exists because of the singularities at $\rho^2=x_{1,2}$  in both equations.

This example allows us to conclude that even if the interaction function $f(\rho)$ satisfies condition~(\ref{fronlex}), the interaction may be too strong for the physical solution to exist. The conclusion most probably applies to a general interaction law~(\ref{key21ph}) as well: repulsive non-linearity should be not too strong to yield meaningful solutions.

Note that a general linear interaction law involving both densities is considered in section~\ref{s6.1}, and two non-linear completely integrable models are found in Appendix~\ref{C}.

\subsection{Non-Singular Cosmological Solutions Starting and Ending 
with Pure Dark Energy (de Sitter Universe)\label{s6}}

Interaction between dark energy and matter allows for cosmological solutions which do not start at a singularity. Instead, their initial and final state is a universe with dark matter only; DE densities in the beginning and end are generally different, and may differ by an arbitrary amount. Below we study such solutions in detail. Note that some non-singular solutions were studied in papers~\cite{Overd98},~\cite{Overd99}, and in~\cite{LimBasSol2}~-~\cite{BasSol} for the $\rv=\rv(H)$ model discussed above .

\subsubsection{A Toy Linear Model: Initial Jump in DE  density\label{s6.1}}

We first explore the general linear interaction law by setting
\be\label{GenLinF}
 F(\rv,\rho)=-s\rho+\theta(\rv-\rinf),\qquad \theta\not=0\; 
\ee
(the case $\theta=0$ is examined in full in section~\ref{s4}). Here $s,\;\theta$ and $\rinf>0$ are the model parameters; the last of them represents the only equilibrium value of uniform DE density possible in this system. It also plays a role of the threshold for self-action of heavy vacuum: if for, say, $\theta<0$, its density is above this critical one, $\rv>\rinf$, then it tries to reduce its amount; in the opposite case $\rv<\rinf$ DE reproduces itself.

According to the expression~(\ref{GenLinF}), the governing equations~(\ref{govsaut1ph}) become:
\be\label{Lineqs}
\frac{d\rho}{d\lambda}=  -\left[(1+w-s)\rho+\theta(\rv-\rinf)\right];\qquad
\frac{d\rv}{d\lambda}= -s\rho+\theta(\rv-\rinf)\; ;
\ee
 as before, $\lambda=\ln\left(V/V_*\right)$. This linear autonomous system has a single equilibrium $\rho=0,\;\rv=\rinf$; we require it to be stable, since we want our solutions to tend exactly to it at large times ($\lambda\to+\infty$). (It is straightforward to see that other cases, when this point is unstable or neutrally stable, do not generically lead to any sound physical solutions.)

So we demand that the  characteristic equation
\be\label{charcteqLin}
\mu^2+(1+w-s-\theta)\mu-(1+w)\theta=0\; 
\ee
of the linear system~(\ref{Lineqs}) with constant coefficients has either a couple of complex conjugate roots with the negative real part, or two negative real roots $\mu_{1}<\mu_{2}<0$,
\be\label{RootsLin}
\mu_{1,2}=0.5\left[-(1+w-s-\theta)\mp\sqrt{\delta}\right],\qquad \delta=(1+w-s-\theta)^2+4\theta(1+w)\; .
\ee
However, in the first case the matter density oscillates around zero and thus does not stay positive all the way, as it should. This leaves us with the second alternative, $\mu_{1}<\mu_{2}<0$, which  condition is guaranteed by the inequalities
\[
\delta>0,\qquad 1+w-s-\theta>0,\qquad (1+w-s)\theta>0 \; .
\]
The analysis shows that they hold for only one range of the parameters, namely (recall that $w\geq0$):
\be\label{ParCondLin}
 s<0,   \qquad \theta<0 \; .
\ee
These inequalities are assumed true in the sequel; note that, due to them, parameter $\theta$ lies between the roots, $\mu_{1}<\theta<\mu_{2}<0$.

The general solution of the linear system~(\ref{Lineqs}) is found in a standard way in terms of exponents of $\lambda$, or, accordingly, powers of $V$:
\bea\label{Mod3LinSol}
\rho=\frac{C_1}{V^{|\mu_{1}|}}+\frac{C_2}{V^{|\mu_{2}|}},\qquad
\rv=\rho_\infty +{|s|}\left(-Q_1\frac{C_1}{V^{|\mu_{1}|}}+Q_2\frac{C_2}{V^{|\mu_{2}|}}\right)\; ;\\
Q_{1}=(\theta - \mu_{1})^{-1}=2\left[(1+w-s+\theta)+\sqrt{\delta}\right]^{-1}>0\; ,\qquad\nonumber\\
Q_{2}=(\mu_{2}-\theta  )^{-1}=2\left[-(1+w-s+\theta)+\sqrt{\delta}\right]^{-1}>0\; ;\qquad\nonumber
\eea
here $C_{1,2}$ are arbitrary constants of integration. For $\rho$ to be positive throughout the expansion  both of them must be positive. But then $\rv$ is negative in the beginning of the expansion, $V\to+0$, since the negative term with $V^{-|\mu_{1}|}$ dominates its expression in this limit. Therefore {\it the general linear interaction law~(\ref{GenLinF}) does not allow for any sound cosmological solutions}.

Still, one special case, $C_{1}<0,\;C_{2}>0$, might turn meaningful with certain addition. In this case we introduce,for convenience, two new constants  $\rst,\Vst>0$:
\[
C_1=-\rst \Vst^{|\mu_{1}|}, \qquad C_2=\rst \Vst^{|\mu_{2}|}\; .
\]
Using these notations, we rewrite the solution~(\ref{Mod3LinSol}) as 
\bea\label{Mod3Lin}
\rho=\rst\left[\left(\frac{\Vst}{V}\right)^{|\mu_{2}|}-\left(\frac{\Vst}{V}\right)^{|\mu_{1}|}\right],\;
\rv=\rho_\infty +{|s|}\rst\left[Q_2\left(\frac{\Vst}{V}\right)^{|\mu_{2}|}+Q_1\left(\frac{\Vst}{V}\right)^{|\mu_{1}|}\right]
\eea

We see that the matter density here goes to negative infinity at the initial moment of time $(V\to+0)$. It stays negative for a finite interval $0<V<\Vst$, becomes zero at $V=\Vst$, and then remains positive for $\Vst<V<+\infty$, vanishing in the limit. In contrast with that, the DE density is always positive, decreasing from positive infinity at $V\to+0$ to $\rinf>0$ in the opposite limit. 

All this is easily seen from the trajectory of solution~(\ref{Mod3Lin}) in the phase plane $\{\rv,\,\rho\}$ plotted in~Fig. 1 (the physical part of this plane is its first quadrant $\rho\geq0,\;\rv\geq0$).  After starting below the horizontal $\rv$ axis, the trajectory crosses it at $V=\Vst$ and stays above it, first going upwards, reaching the highest point, and finally going to the stable equilibrium $\{\rv=\rinf,\,\rho=0\}$ on this axis. Remarkably, the crossing point is at the value of DE density, $\rv(\Vst)$, which is larger than the final value $\rinf$, as implied by the second of the formulas~(\ref{Mod3Lin}):
\[
\rho_0\equiv\rv\Bigl|_{\Vst+0}=\rho_\infty +{|s|}\rst\left(Q_2+Q_1\right)=\rho_\infty +\frac{2|s|\rst\sqrt{\delta}}{(\theta - \mu_{1})(\mu_{2}-\theta  )}>\rinf
\]
\begin{figure}\label{f1}
  \centering
  \includegraphics[width = 14.1 cm]{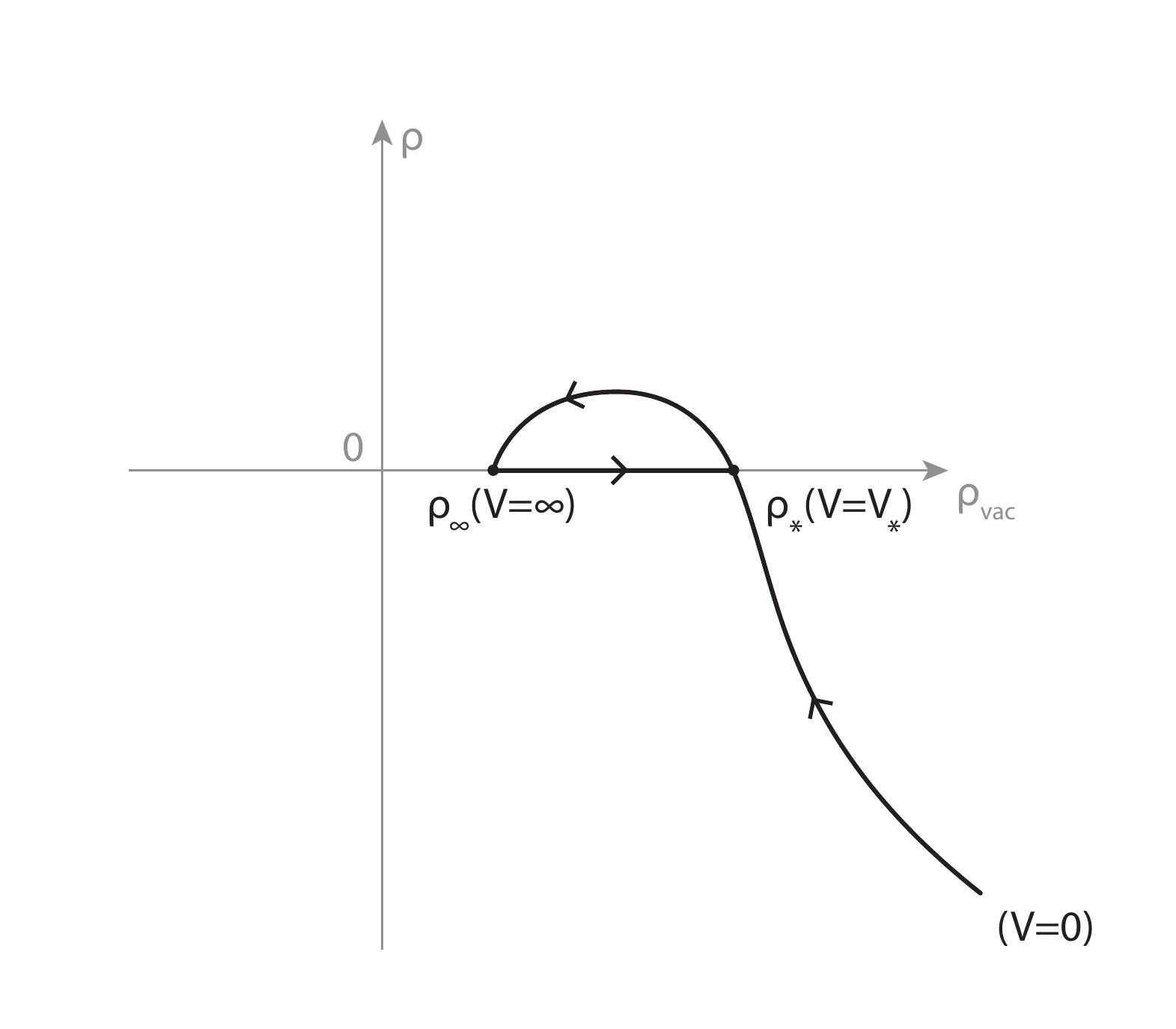}
 \caption{Phase trajectory of a non-physical solution~(\ref{Mod3Lin})  and closed orbit for the solution 
 \centerline{with the initial jump in $\rv$ based on it.}
  }
  \end{figure}

 So, to make this solution more realistic, i.e., the matter density all the way non-negative, one needs just to change it somehow at the initial stretch $0<V<\Vst$ without violating the governing equations. The only available option is to assume that throughout this period of time the system rests at its only equilibrium $\{\rv=\rinf,\,\rho=0\}$, the spacetime is filled with heavy vacuum only, no matter is present. Then at the moment $t_*$ corresponding to $\Vst=V(t_*)$  a positive jump in DE density occurs, driving it up to some value $\rho_0>\rinf$; the corresponding initial conditions, 
  \[
  \rho=0,\qquad \rv=\rho_0\quad\mbox{at}\quad V=\Vst\; ,
  \]
  are then picked up by the governing equations giving the solution~(\ref{Mod3Lin}). Thus the complete solution is:
  \bea\label{Mod3LinPhys}
 \rho=0,\qquad \rv=\rinf,\qquad 0<V<\Vst\; ;\qquad\qquad\qquad\qquad\nonumber\\
\rho=\rst\left[\left(\frac{\Vst}{V}\right)^{|\mu_{2}|}-\left(\frac{\Vst}{V}\right)^{|\mu_{1}|}\right],\;
\rv=\rho_\infty +{|s|}\rst\left[Q_2\left(\frac{\Vst}{V}\right)^{|\mu_{2}|}+Q_1\left(\frac{\Vst}{V}\right)^{|\mu_{1}|}\right],\nonumber\\
\Vst<V<+\infty.\qquad\qquad\qquad\qquad\qquad\qquad\qquad
\eea
It starts and ends with pure heavy vacuum whose initial density is higher - and can be any number of orders of magnitude higher - then the final one. In the phase plane (Fig. 1) this solution corresponds to a finite closed orbit (loop) obtained from the initial infinite one by replacing its part below the horizontal axis with an interval of this axis between $\rinf$ and $\rho_0$, as shown in the figure.

This might represent some interesting physics but for the initial jump in the DE density increasing it instantly. Where does the additional energy come from? Apart from assuming the existence of some other universe(s) connected to the one we are considering, the jump is a clear violation of the energy conservation law. Interestingly, it can be avoided if matter has more than one component leading to a feasible cosmological solution given in sec.~\ref{s7.1}.

Nevertheless, this toy model seems valuable, as it hints to some other ones that do not have the indicated  significant drawback. One can think about multiple rest points and heteroclinic trajectories connecting them; these features, however, belong entirely to  the realm of non-linear models, which we consider next.

\subsubsection{General Non-Linear Model:  Qualitative Picture. Non-Singular Cosmologies Represented by Heteroclynic Phase Trajectories\label{s6.2}}

We now return to the general non-linear case described by the autonomous system~(\ref{govsaut1ph}),
\[
\frac{d\rho}{d\lambda}=  -\left[(1+w)\rho+ F(\rv,\rho)\right];\quad
\frac{d\rv}{d\lambda}= F(\rv,\rho),\qquad  \lambda=\ln\left(V/V_*\right)\; .
\]
Its equilibria and their stability play a central role in what follows, so we first of all recall some related basic facts.

The global stability of a generic rest point of an autonomous system is determined by the local system linearized about this point. The stabilty properties are specified by the behavior of small perturbations of the equilibrium, i.e., by the solutions of this linearized system, whose dependence on the evolution variable $\lambda$ is exponential, $\propto\exp(\mu\lambda)$. The admissible values of the exponent $\mu$ coincide with the set of eigenvalues of the matrix of the linearized system. 

If  each eigenvalue has $\mbox{\bf Re}\,\mu<0$, then small perturbations decay in the vicinity of this rest point, and it is stable (attractive); it is unstable (repulsive) in the opposite case $\mbox{\bf Re}\,\mu>0$ for every eigenvalue, when small perturbations are growing. 

In the case when $\mbox{\bf Re}\,\mu<0$  for a part of eigenvalues, $\mbox{\bf Re}\,k>0$ for another part of them, and perhaps yet  $\mbox{\bf Re}\,\mu=0$ for the remaining third part of them, the equlibrium is called semi-stable. The eignevectors of eignevalues belonging to the first group define the directions of stability, the motion along them goes towards the rest point in its vicinity, i.e., perturbations decay. Accordingly, the second group of eigenvectors define the unstable directions with the motion away from the rest point near it, so the magnitude of perturbations in these directions grows. 

Finally, if all the eigenvalues are purely imaginary, then the rest point is called neutrally stable; such rest points are usually associated with closed orbits near them representing periodic solutions.

Our governing system~(\ref{govsaut1ph}) repeated in the beginning of the current section has the dimension $D=2$, which implies a lot of pleasant specifics (for example, deterministic chaos~\cite{Schultz} possible in all higher dimensions does not occur in the systems on the plane).  The $2\times2$ matrix,  ${\cal M}$, of the system linearized at an equilibrium has just two eignevalues $\kpm$, which are the roots of the quadratic polynomial
\be\label{eqk}
\mu^2-\left(\mbox{tr}\,{\cal M}\right)\mu+\mbox{det}\,{\cal M}=0\; ;
\ee
so they are either real or complex conjugate. 

The rest point is stable if $\mbox{\bf Re}\,\kpm<0$, and unstable in the opposite case $\mbox{\bf Re}\,\kpm<0$. Semi-stable equilibria, called {\it saddles}, correspond to real eigenvalues of the opposite signs, $\km<0,\;\kp>0$. A rest point with imaginary $\kpm$ is called {\it a center}, it is surrounded by closed phase orbits, which  correspond to periodic solutions. 

So the stability condition is $\mbox{det}\,{\cal M}>0,\;\mbox{tr}\,{\cal M}<0$, the instability condition is $\mbox{det}\,{\cal M}>0,\;\mbox{tr}\,{\cal M}>0$. A rest point is a saddle when the discriminant of the polynomial~(\ref{eqk}) is positive and the determinant of the matrix ${\cal M}$ is negative, $\mbox{det}\,{\cal M}<0$; finally, a center occurs when $\mbox{tr}\,{\cal M}=0,\;\mbox{det}\,{\cal M}>0$.

Let now $P_*=\{\rv=\rst,\;\rho=0\}$ be a rest point of our system~(\ref{govsaut1ph}), so that $F(0,\rst)=0$, as in equation~(\ref{DeSit}). The matrix ${\cal M_*={\cal M}(P_*)}$ is then given by
\bea\label{M*}
{\cal M_*}=\left[\matrix{-(1+w+a_*)& -b_*\cr
                 a_* & b_*\cr}\right],\qquad
           a_*=\frac{\partial F}{\partial\rho}\Biggl|_{P_*},    
                               \qquad b_*=\frac{\partial F}{\partial\rv}\Biggl|_{P_*}\; ,\\
															\mbox{tr}\,{\cal M_*}=-(1+w+a_*)+b_*\qquad \mbox{det}\,{\cal M_*}=-(1+w)b_*	\; .\nonumber													
\eea

The stability conditions for $P_*$ are easily calculated to be:
\bea\label{stabcond}
\mbox{saddle:}\;\; b_*>0,\; ;\qquad\qquad\qquad\qquad\quad\nonumber\\
\mbox{stable:}\;\; b_*<0,\qquad b_*<1+w+a_*\; ;\nonumber\\
\mbox{unstable:}\;\; b_*<0,\qquad b_*>1+w+a_* \; ;\\
\mbox{center:}\;\;b_*<0,\qquad b_*=1+w+a_*\; ;\nonumber
\eea
here we took into account $1+w>0$.  In a special case $b_*=0$ perturbations in the $\rho$ direction grow when $a_*>-(1+w)$ and decrease when the opposite inequality holds. However, in the linear approximation perturbations in direction $\rv$ remain constant (corresponding eigenvalue is equal to zero), so stability in this direction should be additionally studied. The same is true for the (non-generic and thus unrealistic) case $b_*=0,\;a_*=-(1+w)$ when both eigenvalues turn to zero.

We are finished with all the preliminaries, and now turn to the general mechanism producing cosmologies dominated by dark energy. We assume that our governing system has (at least) three rest points 
\[
P_j=\{\rv=\rho_j,\;\rho=0\},\; j=1,2,3\; ,
\]
corresponding to three different positive roots of the equation
\[
F(\rho_j,0)=0, \qquad 0<\rho_1<\rho_2<\rho_3\; .
\]
For the reasons that become clear below we will use also the alternative notations,
\[
\rho_1=\rinf,\qquad \rho_2=\rc,\qquad \rho_3=\rz\;.
\]
Moreover, let us assume that the first ($j=1$) and last ($j=3$) rest points are saddles, while the middle one ($j=2$) is a center, or a neutrally stable point, surrounded by closed trajectories representing periodic solutions. Replacing the subscript $_*$ with the subscript $j$ in the stability conditions~(\ref{stabcond}) we write the corresponding requirements as
\be\label{sadcensad}
b_{1,3}>0;\qquad b_2<0,\quad b_2=1+w+a_2\;.
\ee

Under these conditions, easily met, of course, by a general function $F(\rv,\rho)$, heteroclinic trajectories, or separatrices, that go from one saddle point to the other, {\it necessarily exist}. They separate the finite closed phase orbits around the center $\{\rho=0,\,\rv=\rho_c\}$ from the infinite ones (this statement is valid only in the plane, $D=2$!). Each separatrix starts from one saddle along its unstable direction, and ends at the other one, approaching it along its stable direction.

In this way, a classical phase portrait of the system appears that is given in Fig.~2 in solid lines. The arrows on heteroclinic curves show the direction of motion when the evolution variable ($\lambda$, or $V$, or $t$) increases; it takes an infinite time to go from one end of the separatrix to the other.
\begin{figure}\label{f2}
  \centering
  \includegraphics[width = 14.1 cm]{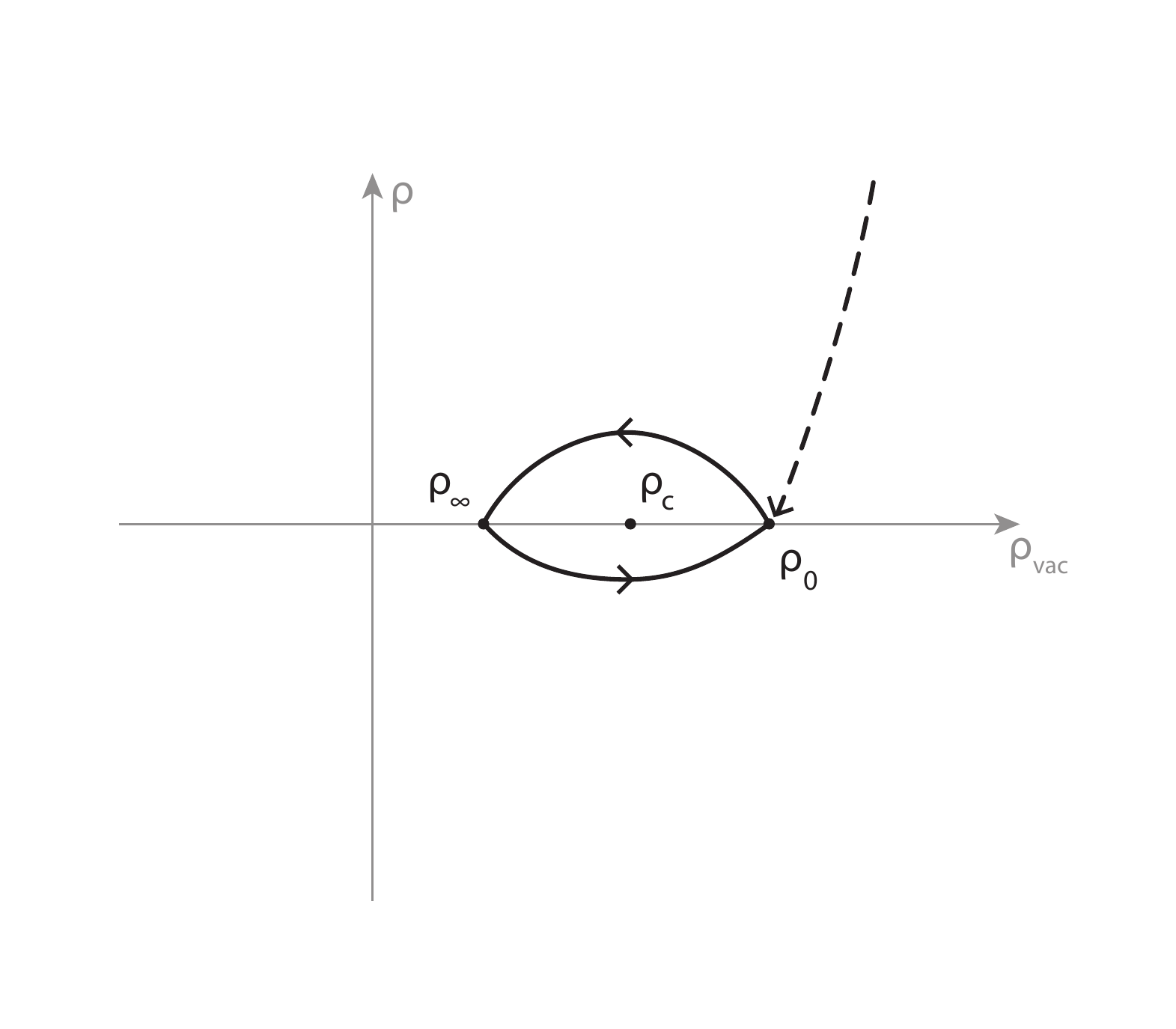}
 \caption{Two saddles connected by heteroclinic curves surrounding a single center.
 The upper curve is a trajectory corresponding to a non-singular cosmological solution starting and ending with pure dark energy of different densities. Dashed line is a trajectory corresponding to some solution starting at a singularity.}
    \end{figure}

The upper  heteroclinic curve corresponds, in fact, to a valid non-singular cosmological solution that starts with a pure heavy vacuum of the density $\rho_0$, and ends again in the state with no matter and DE of a smaller density $\rinf$. Matter appears from the vacuum due to their interaction, its density grows and reaches some maximum $\rho=\rho_{max}$, and then decreases to zero when $t\to+\infty$.
\begin{figure}\label{f3}
  \centering
  \includegraphics[width = 14.1 cm]{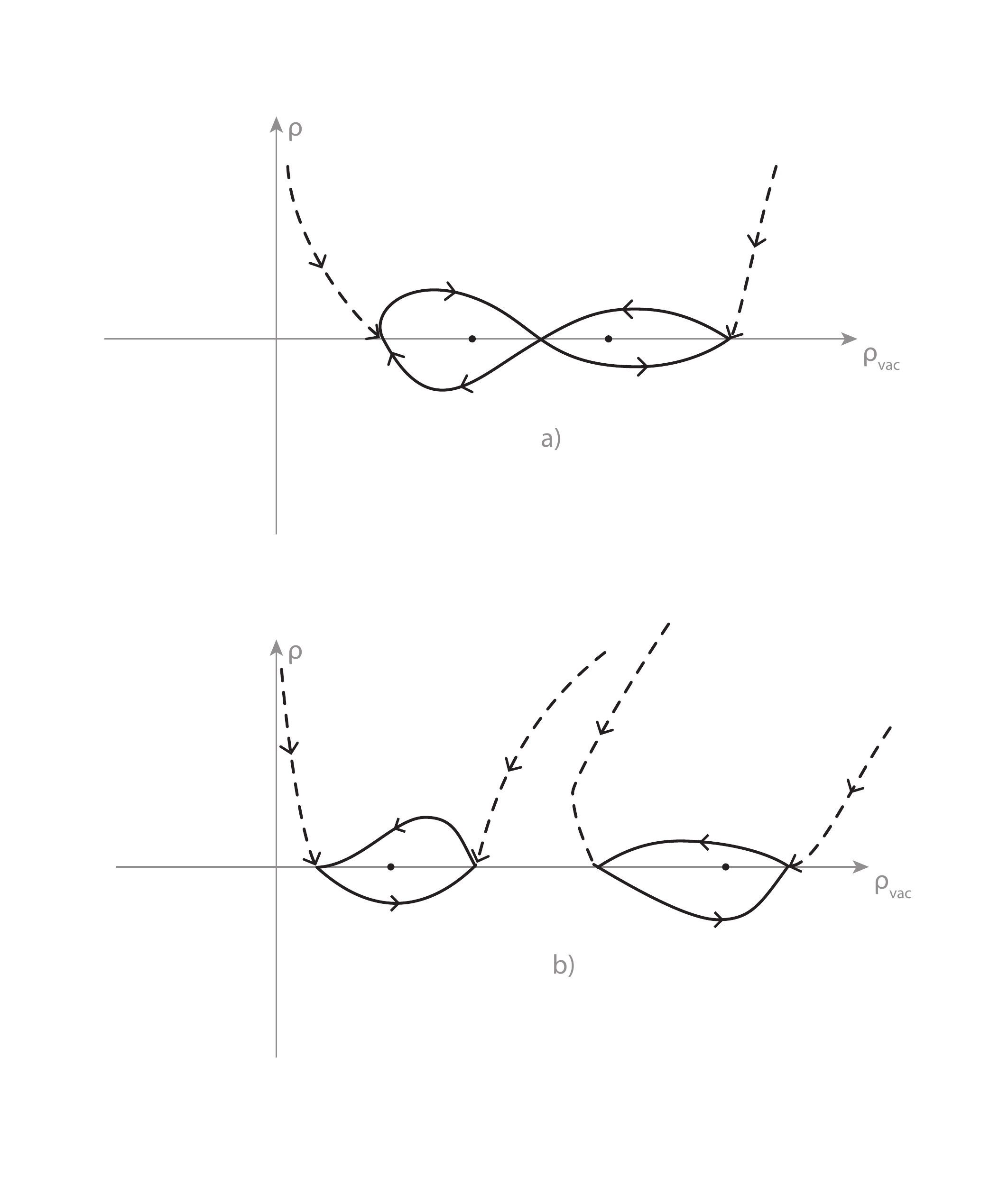}
 \caption{More saddles (crosses), centers (dots), and heteroclinic curves surrounding the latter: a) - 3 saddles, 2 centers (`cat's eyes'); b)- 4 saddles, 2 centers.  Each positive heteroclinic trajectory corresponds to a non-singular cosmological solution starting and ending with pure DE. Dashed lines are trajectories of singular cosmological solutions.}
    \end{figure}

This solution can be realized in the following way. Initially, the whole spacetime rests at the de Sitter equilibrium $P_3$, it is filled with DE of the density $\rz$ and nothing else. At some moment $t=0\;(V=0)$ due to a small perturbation of this equilibrium in its unstable direction toward positive values of $\rho$, the universe gets off $P_3$ to the upper separatrix, and goes along it to another de Sitter equilibrium $P_1$ at $t=+\infty$. Such instability can happen for various physical reasons, for example, it may occur due to the particle creation~\cite{Mot}.

Of course, depending on the intricacies of the DE--matter interaction, there can be more saddles and more heteroclinic trajectories representing non-singular cosmological solutions (examples with 3 and 4 saddles are given in Fig. 3).
Moreover, a heteroclinic trajectory might go not to a saddle, but to a  stable rest point from either an unstable one, or a saddle. In the former case of an unstable and stable equilibria {\it an infinite set} of heteroclinic curves can exist, as in Fig. 4a.
\begin{figure}\label{f4}
  \centering
  \includegraphics[width = 14.1 cm ]{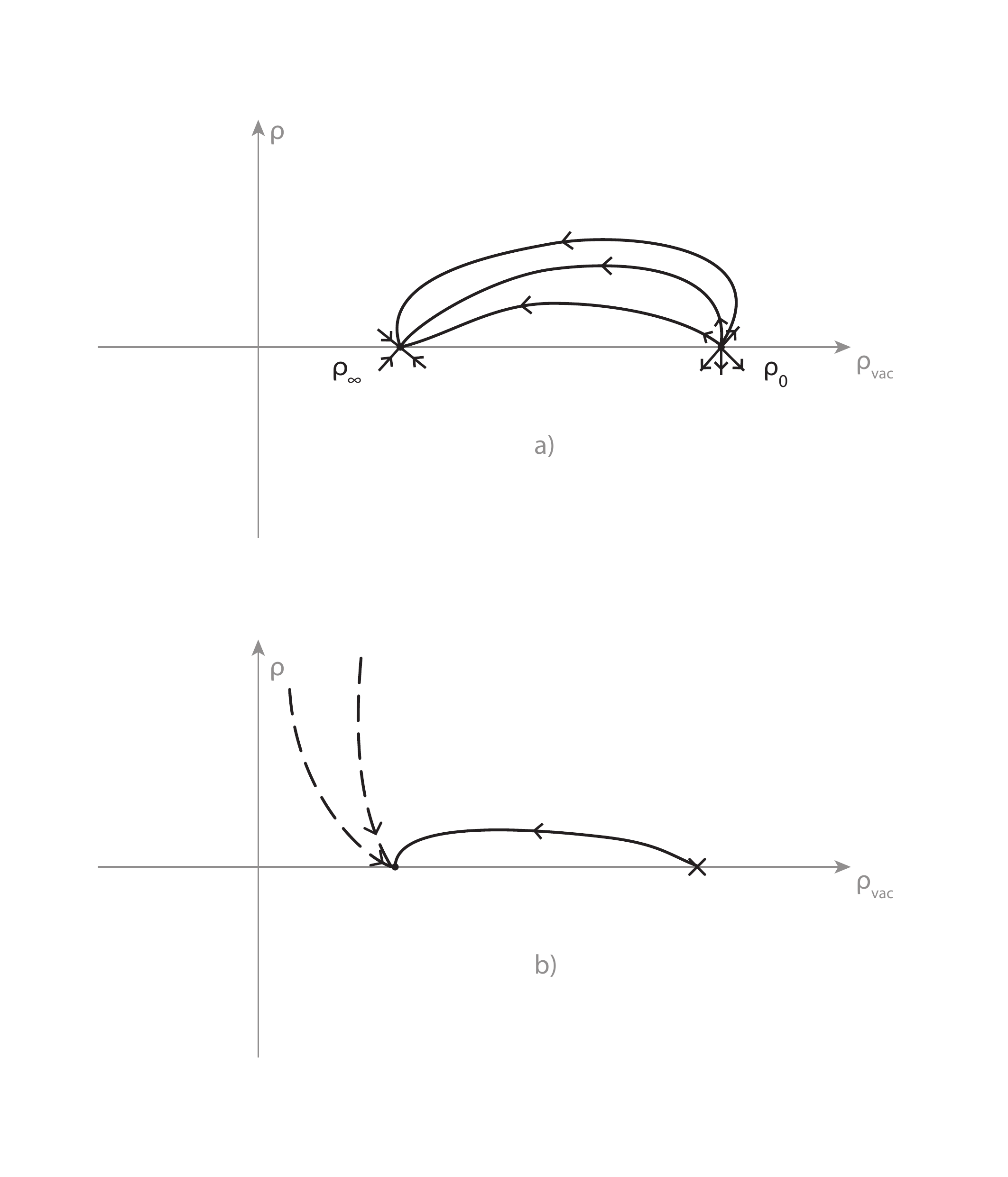}
 \caption{a) - trajectories of non-singular cosmological solutions starting at an unstable and ending at the stable de Sitter equilibrium. b) -  trajectory of a non-singular cosmological solution starting at a saddle and ending at the stable de Sitter equilibrium. Dashed lines are trajectories of singular cosmological solutions.}
    \end{figure}

Independent of how it occurs, {\it each positive heteroclinc trajectory corresponds to a non-singular cosmological solution starting and ending at the de Sitter equilibria with diffrent DE densities}. There also can be {\it homoclinic} curves going from a saddle back to the same saddle, in which, cleraly non-generic, case the initial density of heavy vacuum is equal to the final one. In any case, all non-singular cosmologies start and end at pure vacuum states, because there are only equilibria with $\rho=0$.

In addition, `standard' cosmologies that emerge from singularities and correspond to infinite phase trajectories tending to a rest point at large times, such  as those drawn in dashed lines in Figs. 2 and 3, are usually present. This infinite variety of cosmological solutions represents a `multi-verse' that can be created by a common heavy vacuum due to its complex interaction with matter.

Significant features are unveiled by studying the time dependence of the non-singular cosmological solutions, described by heteroclinic trajectories, at there beginning and end. As usual, this time dependence is derived by asymptotically calculating the integral in the basic relation (\ref{volt}). Since $\rho\to+0,\;\rv\to\rinf$ at large times, the asymptotics of the scale factor is an exponential one, exactly as in (\ref{Mod1tlarg}):
\be\label{hettlarg}
 a(t)\sim\exp(t/\tau_\infty),\qquad  \tau_\infty= \sqrt{8\pi\rho_\infty/3}\; .
 \ee

The situation with the behavior at small times is more complicated. For the open universe, $k=-1$,  the curvature term $V^{-2/3}$ is positive and dominating, in the limit  $V\to+0$ under the square root in the eq.~(\ref{volt}). So a non-singular solution can exist that starts from the zero value of the scale factor, or co--moving volume:
\be\label{hettsmall}
V^{2/3}(t)\sim t,\qquad a(t)\sim \sqrt{t}, \qquad t\to+0\; .
\ee

If the universe is closed, $k=1$, then the expression under the square root  in the formula~(\ref{volt}) becomes negative for small values of $V$ due to the negative curvature contribution $-V^{-2/3}$. Hence there is no meaningful non-singular solution describing a closed universe that starts at $a=0$. 

In the case of a flat universe, $k=0$, the total density tends to $\rho_0$ at the expansion beginning, which results in
\[
a(t)= a_0\,\exp(t/\tau_0),\qquad \tau_0= \sqrt{8\pi\rho_0/3}\; ,
\]
so $a\to+0$ only when $t\to-\infty$; the solution is defined on the whole time axis. 

However, unlike the singular case, there is no need for a non-singular cosmological solution to start at the zero value of the scale factor: before the expansion starts, a static de Sitter universe exists whose scale factor grows exponentially and can have any positive value at any given moment of time. Therefore an alternative for a universe of any curvature is to start expanding, at $t=0$, with a finite scale factor $a_0=a(0)>0$. 

For the closed universe the minimum starting scale factor value is defined by the initial DE density, $\rz$:
\[
a_0=(8\pi\rz)^{-1/2}\; 
\]
The initial velocity of expansion is then equal to zero, $\dot{a}(0)=0$, as implied by the first  Friedmann equation~(\ref{FrEq}). The scale factor is a regular function of time near $t=0$, its two--term Taylor expansion is
\be\label{a0zerodota0}
a(t)=a_0\left(1+t^2/6a_0^2+\dots\right), \;\; t\to+0;\qquad a_0=(8\pi\rz)^{-1/2},\;\; k=1\; .
\ee

A closed universe can also start at any scale factor value larger than the minimum one, with the corresponding finite velocity. A flat or open non-singular  universe can start at any positive value of the scale factor. In all these cases the scale factor is regular at $t=0$, with the following two--term expansion: 
\be\label{a0zero}
a(t)=a_0\left(1+t/\tau+\dots\right), \;\; t\to+0;\quad \tau=\sqrt{3/\left(8\pi\rz-k/a_0^2\right)},\quad k=0,\pm1\; .
\ee

\subsection{Non-Singular Cosmologies:  Exact Solutions\label{s6.3}}

\subsubsection{General  Exact Solution by the Semi--Inverse Method\label{s6.3.1}}

Explicit construction of heteroclinic solutions is always difficult, even if the governing equations are explicitly integrable, which is not the case of our system~(\ref{govsaut1ph}) with a general interaction law $F(\rv,\rho)$. Luckily, a semi--inverse solution method comes to rescue.

We assume that our system has at least two rest points, $P_0=\{\rho_0,\;0\}$ and $P_\infty=\{\rho_\infty,\;0\}$, $\rho_\infty<\rho_0$, and that there exists a positive heteroclinic phase trajectory ${\cal H}$ connecting the first point with the second one, as in Figs. 2---4. We are looking for the exact solution describing this trajectory. Along it, the matter density is some smooth enough positive function, $h(\rv)$, of the DE density,  which turns to zero at both ends of the interval $\rho_\infty<\rv<\rho_0$:
\be\label{rhoh}
\rho=h(\rv);\qquad h(\rv)>0\;\;\mbox{for}\;\; \rho_\infty<\rv<\rho_0;\qquad h(\rho_\infty)=h(\rho_0)=0\; .
\ee

Our plan is to keep $h(\rv)$ otherwise arbitrary, and try to find the proper expression for the interaction function $F(\rv,\rho)$ along the heteroclinic curve that makes the function~(\ref{rhoh}) to satisfy the first of the  governing equations~(\ref{govsaut1ph}). We then try to complete the solution by finding some meaningful $\rv(V)$ from the second of them. This is what we call the semi--inverse method; if successful, it allows one to obtain a non-singular cosmological solution and study its properties.

At the first step of our approach, we combine the second of the equations~(\ref{govsaut1ph}),
\[
\frac{d\rv}{d\lambda}= F(\rv,\rho)\; ,
\]
with the representation~(\ref{rhoh}) on the heteroclinic curve to get:
\[
\frac{d\rho}{d\lambda}\biggl|_{{\cal H}}=\frac{dh}{d\rv}\,\frac{d\rv}{d\lambda}\biggl|_{{\cal H}}=h^{'}(\rv)F(\rv)=h^{'}(\rv)F(\rv,h(\rv))\; .
\]
Therefore the first of the governing equations,
\[
\frac{d\rho}{d\lambda}=  -\left[(1+w)\rho+ F(\rv,,\rho)\right]\; ,
\]
turns, along the curve ${\cal H}$, into
\[
h^{'}(\rv)F(\rv,h(\rv))=-\left[(1+w)h(\rv)+F(\rv,h(\rv))\right]\; ,
\]
giving thus
\be\label{Fh}
F\biggl|_{{\cal H}}=F(\rv,h(\rv))=-(1+w)\frac{h(\rv)}{1+h^{'}(\rv)}\; .
\ee
This is the result of the first step of our approach; it requires two comments.

First, expression~(\ref{Fh}) specifies the interaction function $F(\rv,\rho)$ {\it on the heteroclinic curve $\rho=h(\rv)$  only}. Its value in the rest of the physical quarter--plane $\rv\geq0,\;\rho\geq0$ can be provided by an infinite number of smooth (one time continuously differentiable) extensions of ~(\ref{Fh}). The most obvious extension is 
\[
F(\rv,\rho)=-\frac{(1+w)\rho}{1+f(\rv,\rho)};\qquad f(\rv, h(\rv))=h^{'}(\rv)\; .
\]
However, it represents a `degenerate' case: each point of the positive semi--axis $\rv\geq0$ is a rest point here. To avoid this, one can use other extensions, like
\bea
F(\rv,\rho)=-\frac{(1+w)\rho+f_1(\rv,\rho)}{1+f_2(\rv,\rho)}\; ,\qquad\qquad\nonumber\\
f_1(\rv, h(\rv))=0,\qquad f_2(\rv, h(\rv))=h^{'}(\rv)\; ,\nonumber
\eea
and so on. Depending on the extension, or, better to say, on the complete law of DE--matter interaction, the system may or may not have rest points other than $P_0$ and $P_\infty$. Moreover, the stability of the rest points also  depends on the extension of the expression~(\ref{Fh}); however, the heteroclinic curve connects $P_0$ with $P_\infty$, so $P_0$ is either a saddle or an unstable equilibrium, while $P_\infty$ is a stable one or a saddle, as in Fig. 4.

The second comment is that we have to avoid singularities of the function~(\ref{Fh}) only within the interval $\rho_\infty<\rv<\rho_0$: all other singularities can be eliminated by choosing the extension appropriately. If exist, the singularities of $F\bigl|_{{\cal H}}$  are the zeros of the denominator in~(\ref{Fh}); since $h^{'}(\rv)$ is positive near the left end of the interval and negative at the right one, there are no such zeros if and only if
\be\label{nosing}
\min_{\rho_\infty\leq\rv\leq\rho_0}h^{'}(\rv)>-1\; .
\ee
This is a regularity condition for the function~(\ref{Fh}), and simultaneously one more restriction on the function $h(\rv)$.

At the next step of the solution by the semi--inverse method we integrate the second governing equation along the heteroclinic curve, where, by the formula~(\ref{Fh}), 
\[
\frac{d\rv}{d\lambda}=-(1+w)\frac{h(\rv)}{1+h^{'}(\rv)}\; .
\]
 The result of this simple integration in terms of the variable $V$ is:
\be\label{rvac}
h(\rv)\exp H(\rv)=\rst\left(\frac{\Vst}{V}\right)^{1+w},\qquad H(\rv)=\int\limits^{\rv}\,\frac{dv}{h(v)}\; .
\ee
An arbitrary constant $\rst>0$ is introduced for the consistency of writing; of course, effectively there is only one arbitrary constant here, that is,
\[
C_*=\rst\Vst^{1+w}\; .
\]
If, for a  given $h(\rv)$, the equation~(\ref{rvac}) has a proper solution $\rv(V)$, particularly, with the right behavior at the ends of the interval, $\rv\to\rinf+0$  and $\rv\to\rho_0-0$, then, combined with  $\rho=h(\rv(V))$, it provides a valid cosmological solution.  The existence of such solution can hardly be established in general; instead, one can usually successfully analyze equation~(\ref{rvac}) for  a particular $h(\rv)$. 

However, assuming that the solution does exist, we verify its limit behavior in Appendix~\ref{A} under the assumption that the zeros of $h(\rv)$ at $\rv=\rho_0,\rho_\infty$ are general algebraic, i.e.:
\bea\label{asymph}
h(\rv)=h_\infty(\rv-\rho_\infty)^{\nu_\infty}[1+o(1))],\;\;\rv\to\rinf+0; \quad h_\infty>0,\;\; \nu_\infty>0\; ;\nonumber\\
h(\rv)=h_0(\rho_0-\rv)^{\nu_0}[1+o(1))],\;\;\rv\to\rho_0-0\quad\; h_0>0,\;\; \nu_0>0\; .\;\;
\eea
Assuming also that these asymptotic expressions can be differentiated in $\rv$,
\bea\label{asymphprime}
h^{'}(\rv)=\nu_\infty h_\infty(\rv-\rho_\infty)^{\nu_\infty-1}[1+o(1))],\quad\rv\to\rinf+0\; ;\qquad\quad\;\;\nonumber\\
h^{'}(\rv)=-\,\nu_0 h_0(\rho_0-\rv)^{\nu_0-1}[1+o(1))],\quad\rv\to\rho_0-0\; ,\qquad\qquad
\eea
we conclude that
\bea
h^{'}(\rho_0)=0,\;\;\nu_0>1;\qquad h^{'}(\rho_0)=-h_0,\;\;\nu_0=1;\qquad h^{'}(\rz)=-\infty,\;\;0<\nu_0<1\;;\quad\nonumber\\
h^{'}(\rho_\infty)=0,\;\;\nu_\infty>1;\quad h^{'}(\rho_\infty)=h_\infty,\;\;\nu_\infty=1;\quad h^{'}(\rho_\infty)=+\infty,\;\;0<\nu_\infty<1\; .\quad\;
\nonumber 
\eea
This shows  that the condition~(\ref{nosing}) guaranteeing the lack of singularity of the interaction function on the heteroclinic curve results in the following restrictions on the parameters:
\be\label{nu0}
\nu_0>1;\qquad \nu_0=1,\quad 0<h_0<1\; .
\ee
So, the zero of $h(\rv)$ at $\rv=\rho_0$ should be of order one at least; the positive constants $h_\infty$ and $\nu_\infty$ remain unrestricted. 

The asymptotic analysis of Appendix~\ref{A}  demonstrates that, under the conditions~(\ref{asymph}) -  (\ref{nu0}),  the non-singular solutions obtained by the semi--inverse method can start only with the zero value of the scale factor. According to the previous section, this means that such solutions  describe only an open universe. (However, there is no such  limitation in the case of several matter components when not all of them interacting with dark energy, see section~\ref{s7}).

Combining the results of this section with those of Appendix~\ref{A} we arrive at the following description of the  solutions found by the semi--inverse method.

{\it Let $P_0=\{\rho_0,\;0\}$ and $P_\infty=\{\rho_\infty,\;0\}$, $\rinf<\rho_0$, be the rest points of the system}~(\ref{govsaut1ph}) {\it governing cosmological evolution. Let a smooth enough function $h(\rv)$ be positive in the interval $\rinf<\rv<\rho_0$, turn to zero at both its ends, and satisfy condition} (\ref{nosing}){\it .  Let the function $F(\rv,\rho)$ describing the DE--matter interaction be restricted to the heteroclinic curve $\rho=h(\rv)$ connecting $P_0$ and $P_\infty$ according to the equation}~(\ref{Fh}).

{\it If equation}~(\ref{rvac})  {\it has a continuously differentiable positive solution  $\rv=\rv(V)$ such that  $\rv(+0)=\rho_0,\;\rv(+\infty)=\rinf$, then
\[
\rho=h(\rv)=h(\rv(V)),\qquad \rv=\rv(V)
\]
is a solution of the governing system with the above heteroclinic curve as its phase trajectory. It describes a Friedmann universe whose expansion starts with pure dark energy  of the density $\rho_0$ and no singularity, and ends with dark energy of the density $\rho_\infty<\rho_0$. 

If, in addition,  $h(\rv)$ satisfies conditions}~(\ref{asymph}) {-} (\ref{nu0}), {\it then the asymptotic dependence of both densities on time is:}
\vfill\eject
\bea\label{ast0}
\underline{t\to+0,\quad a(t)\sim\sqrt{t}\;\;(\mbox{open universe})}
\qquad\qquad\qquad\qquad\quad\qquad\qquad\quad\nonumber\\
a)\;\mbox{for}\quad \nu_0>1,\qquad\qquad\quad\qquad\qquad\quad\qquad\qquad\quad\qquad\qquad\quad\qquad\qquad\quad\qquad\qquad\quad\qquad\nonumber\\
\rho_0-\rv\sim\left(\ln t\right)^{-\frac{1}{\nu_0-1}},\qquad\qquad\qquad \rho\sim\left(\ln t\right)^{-\frac{\nu_0}{\nu_0-1}}\;;\qquad\;\;\qquad\;\;\qquad\qquad\nonumber\\
b)\;\mbox{for}\quad \nu_0=1,\;\;0<h_0<1,
\qquad\qquad\quad\qquad\qquad\quad\qquad\qquad\quad\qquad\qquad\quad\qquad\qquad\qquad\\
\rho_0-\rv\sim\rho\sim t^{\frac{3(1+w)h_0}{2(1-h_0)}}\; .\qquad\qquad\quad\qquad
\qquad\qquad\;\;\qquad\qquad\quad\qquad\qquad\nonumber
\eea
\bea\label{astinf}
\underline{
t\to+\infty,\quad a(t)\sim\exp(t/\tau_\infty),\quad\tau_\infty=\sqrt{8\pi\rho_\infty/3}\;\;(\mbox{open universe}) 
}\qquad\quad\qquad\qquad\qquad\nonumber\\
a)\;\mbox{for}\quad \nu_\infty>1,\qquad\qquad\quad\qquad\qquad\quad\qquad\qquad\quad\qquad\qquad\quad\qquad\qquad\quad\qquad\qquad\quad\qquad\nonumber\\
\rv-\rho_\infty\sim t^{-\frac{1}{\nu_\infty-1}},\qquad\qquad\qquad\qquad
\rho\sim t^{-\frac{\nu_\infty}{\nu_\infty-1}}\;;\qquad\qquad\qquad\quad\qquad\quad\nonumber\\
b)\;\mbox{for}\quad \nu_\infty=1,\qquad\qquad\;\,
\qquad\qquad\quad\qquad\qquad\quad\qquad\qquad\quad\qquad\qquad\quad\qquad\qquad\qquad\quad\;\;\\
\rv-\rho_\infty\sim\rho\sim\exp\left[-\frac{3(1+w)h_\infty}{2(1+h_\infty)}\frac{t}{\tau_\infty}\right]\; ;\quad\qquad\qquad\qquad\qquad\quad
\quad\qquad\quad\nonumber\\
c)\;\mbox{for}\quad \nu_\infty<1,\qquad\qquad\quad\qquad\qquad\quad\qquad\qquad\quad\qquad\qquad\quad\qquad\qquad\quad\qquad\qquad\quad\qquad\nonumber\\
\rv-\rho_\infty\sim\exp\left[-\frac{3(1+w)}{2\nu_\infty}\frac{t}{\tau_\infty}\right],\qquad\qquad\quad
\rho\sim\exp\left[-\frac{3(1+w)}{2}\frac{t}{\tau_\infty}\right]\; .\qquad\qquad\quad\nonumber
\eea

Remarkably, for $\nu_0>1$, when $\rho_0$ is a zero of $h(\rv)$ of the order higher than one, both densities evolve very slowly in the beginning, only as an inverse power of the log of time. Similarly, at large times they both tend to their limits only as inverse powers for $\nu_\infty>1$,  with the exponential decay in other cases.

\subsubsection{Particular Non--Singular Cosmologies: Examples of Exact Solutions\label{s6.3.2}}

Any particular choice of the function  $h(\rv)$~(\ref{rhoh}) satisfying all the pertinent conditions  and allowing for the proper solution $\rv(V)$ of the equation~(\ref{rvac}) provides an example of the solution describing some non--singular cosmology. We are choosing simple enough expressions for $h(\rv)$, in particular, allowing for the integral $H(\rv)$ to be calculated in terms of elementary functions.
\vskip1mm
\noindent\underline{Example 1} Clearly, the simplest possible expression for  $h(\rv)$ is
\be\label{h1}
h(\rv)=\frac{(\rz-\rv)(\rv-\rinf)}{R},\qquad R>0\; .
\ee
It satisfies conditions~(\ref{rhoh}),~(\ref{asymph}), and~(\ref{asymphprime}) with  parameter values
\be\label{nuh1}
\nu_0=\nu_{\infty}=1,\qquad h_0=h_{\infty}=(\rz-\rinf)/{R}\; .
\ee
The inequality~(\ref{nu0}) on the value of $h_0$  is met when
\be\label{par1}
h_0=(\rz-\rinf)/{R}<1, \quad\mbox{or}\quad R>\rz-\rinf\; ;
\ee
it also ensures the inequality~(\ref{nosing}) required to avoid singularities of the interaction function. Therefore the function~(\ref{h1}) with the parameters restricted by the inequality~(\ref{par1}) meets all the desirable conditions.

It is straightforward to calculate the integral~(\ref{rvac}) for this case, with the result:
\be\label{H1}
H(\rv)=\ln\left(\frac{\rv-\rinf}{\rz-\rv}\right)^{1/h_0}\; ;
\ee
thus the equation~(\ref{rvac}) for $\rv(V)$ reduces to
\be\label{rvV1}
(\rv-\rinf)^{(1/h_0)+1}=\frac{K}{V^{1+w}}(\rz-\rv)^{(1/h_0)-1},
\qquad K=R\rst\Vst ^{1+w}=RC_{*}>0\; .
\ee
Since, by~(\ref{par1}), $1/h_0>1$, the l.h.s. of this equation {\it increases} monotonically from zero at $\rv=\rinf$ to a positive value at $\rv=\rz$. Contrary to this, the r.h.s. monotonically {\it decreases} to zero at  $\rv=\rz$, for any $V>0$. Therefore the two curves have a single intersection, i.e., the equation has the unique solution $\rv(V),\;\;0<V<\infty$. 

So, equations~(\ref{h1}) and~(\ref{rvV1}) under the condition~(\ref{par1})  define a solution of the governing equations corresponding to a non-singular cosmology. Its behavior in the beginning and end of the expansion is described by the formulas~(\ref{ast0}) and~(\ref{astinf}):
\bea\label{as1}
\rho_0-\rv\sim\rho\sim t^\alpha,\quad \alpha=3(1+w)(\rz-\rinf)/2\left[R-(\rz-\rinf)\right],\quad t\to+0\; ;\nonumber\quad\\
\rv-\rho_\infty\sim\rho\sim\exp
\left\{
-\frac{3(1+w)(\rz-\rinf)}{2[R-(\rz-\rinf)]}\frac{t}{\tau_\infty}
\right\},
\;\; \tau_\infty=\sqrt{8\pi\rho_\infty/3}; \quad t\to+\infty\; .\nonumber
\eea

In this example the heteroclinic phase trajectory~(\ref{h1}) is the top of a quadratic parabola with its maximum at the midpoint, $(\rz+\rinf)/2$, of the interval. Thus the maximum value of matter density achieved in the course of evolution is
\be\label{rmax1}
\rho_{max}=h\left(\frac{\rz+\rinf}{2}\right)=\frac{(\rz-\rinf)^2}{4R}<\frac{\rz-\rinf}{4} \; ;
\ee
the inequality here is implied by the condition~(\ref{par1}). Naturally, $\rho_{max}$ tends to zero when $R\to\infty$ and $(\rz-\rinf)$ is fixed.
\vskip1mm
\noindent\underline{Example 2}

Formula~(\ref{h1}) can be generalized to
\[
h(\rv)={(\rz-\rv)(\rv-\rinf)}q(\rv)\; ,
\]
where the function $q(\rv)$ is positive on the segment $\rinf\leq\rv\leq\rz$ and has a bounded derivative. As our next example, we take $q(\rv)=\theta/\rv$, $\theta>0$, so that
\bea\label{h2}
h(\rv)=\theta(\rz-\rv)(\rv-\rinf)/\rv\; ;\qquad\qquad\qquad\\
\nu_0=\nu_{\infty}=1,\qquad h_0=\theta(\rz-\rinf)/\rz,\quad h_\infty=\theta(\rz-\rinf)/{\rinf}\; .\nonumber
\eea
Condition~(\ref{par1}), $h_0<1$, requires
\[
\theta<\rz/(\rz-\rinf)\; .
\]
Since
\be\label{h'2}
h^{'}(\rv)=\theta(\rz\rinf-\rv^2)/\rv^2\; ,
\ee
the inequality~(\ref{nosing}), $h^{'}(\rv)>-1$, holds for
\be\label{par2}
0<\theta<(\rz+\rv)/\rz\; ,
\ee
which also guarantees the previous inequality, because
$(\rz+\rinf)/\rz<\rz/(\rz-\rinf)$. 

Next we calculate $H(\rv)$ (compare to the expression~(\ref{H1})):
\[
H(\rv)=\ln\frac{(\rv-\rinf)^{1/h_{\infty}}}{(\rz-\rv)^{1/h_{0}}}\; .
\]
Thus the equation~(\ref{rvac}) for $\rv(V)$ is:
\be\label{rvV2}
(\rv-\rinf)^{(1/h_0)+1}=\frac{K}{V^{1+w}}\rv(\rz-\rv)^{(1/h_0)-1},
\quad K=\theta^{-1}\rst\Vst ^{1+w}>0\; .
\ee
It differs from the equation~(\ref{rvV1}) of Example 1 by a single {\it growing} factor $\rv$ on the utmost right, which however causes the additional investigation of how the r.h.s. behaves as a function of $\rv$, since its second factor is decreasing. The analysis shows that the r.h.s., i.e., the product of the two factors, is decreasing under the condition $\theta<\rz/(\rz-\rinf)$, which is true by virtue of the condition~(\ref{par2}). Therefore  the equation~(\ref{rvV2}) has a unique positive solution $\rv(V)$, as in the Example 1. 

So, a unique non-singular cosmological solution exists in this case for any parameters satisfying condition~(\ref{par2}). Its small and large time behavior is described by the following asymptotic formulas,
\bea\label{as2}
\rho_0-\rv\sim\rho\sim t^\alpha,\quad \alpha=3(1+w)\theta(\rz-\rinf)/2\left[\rz-\theta(\rz-\rinf)\right]\,\quad t\to+0\; ;\nonumber\;\;\\
\rv-\rho_\infty\sim\rho\sim\exp
\left\{
-\frac{3(1+w)\theta(\rz-\rinf)}{2[\rz-\theta(\rz-\rinf)]}\frac{t}{\tau_\infty}
\right\},
\;\; \tau_\infty=\sqrt{8\pi\rho_\infty/3}; \quad t\to+\infty\; ,\nonumber
\eea
which can be obtained from the formulas~(\ref{as1}) of Example 1 by a formal replacement of $R$ with $\rz/\theta$.
The main difference between the solutions from the two examples is that the phase trajectory of the second one is no longer symmetric about the midpoint $(\rinf+\rz)/2$. In particular, the maximum value of the matter density is achieved when $\rv=\sqrt{\rz\rinf}$:
\be\label{rmax2}
\rho_{max}=h(\sqrt{\rz\rinf})=\theta(\sqrt\rz-\sqrt\rinf)^2\; .
\ee
\vskip1mm
\noindent\underline{Example 3}

In the previous two examples the zeros of the function $h(\rv)$ at both ends of the interval were of the first order. In other words, the phase trajectory $\rho=h(\rv)$ intersected the $\rv$ axis at a non-zero angle at both $\rv=\rinf$ and $\rv=\rz$. Now we consider an example  of higher, second order zero, when the phase orbit just touches the horizontal axis. We assume
\be\label{h3}
h(\rv)={(\rz-\rv)(\rv-\rinf)^2}/{R^2},\qquad R>0\; ,
\ee
so
\be\label{nuh3}
\nu_0=1,\quad h_0=(\rz-\rinf)^2/{R^2};\qquad \nu_{\infty}=2,\quad h_{\infty}=(\rz-\rinf)/{R^2}\; .
\ee
Conditions (\ref{nu0}) and (\ref{nosing}) are met by just
\be\label{h03}
R>\rz-\rinf\; .
\ee
The function $H(\rv)$ is
\[
H(\rv)=\ln\left(\frac{\rv-\rinf}{\rz-\rv}\right)^{1/h_0}-\frac{1}{h_0}\,\frac{\rz-\rinf}{\rv-\rinf} \; .
\]
 Therefore the key equation (\ref{rvac}) turns into
\be\label{rvV3}
(\rz-\rv)^{(1/h_0)+2} \exp{\left(-\frac{1}{h_0}\,\frac{\rz-\rinf}{\rv-\rinf}\right)}=\frac{K}{V^{1+w}}(\rv-\rinf)^{(1/h_0)-1}\; ,
\ee
where $K=R^2\rst\Vst ^{1+w}>0$. Since $1/h_0>1$, the same arguments of monotonicity apply here as in the two previous examples, so the unique positive solution $\rv=\rv(V)$ to the equation (\ref{rvV3}) exists for any $V>0$. Along with $\rho=h(\rv)$, it defines yet another non-singular cosmology; its behavior at the beginning and end of the expansion is described by the formulas:
\bea\label{as3}
\rho_0-\rv\sim\rho\sim t^\alpha,\quad \alpha=3(1+w)(\rz-\rinf)^2/2\left[R^2-(\rz-\rinf)^2\right],\quad t\to+0\; ;\nonumber\quad\\
\rv-\rho_\infty\sim t^{-1},\qquad \rho\sim t^{-2}, \quad t\to+\infty\; .\qquad\qquad\qquad\qquad\nonumber
\eea
Remarkably, the behavior of the density at large times is no longer exponential but given by a power law; this is always true for $\nu_\infty>1$, according to the formula a) in~(\ref{astinf}).

The maximum value of the matter density in the course of the expansion is
\be\label{rmax3}
\rho_{max}=h\left(\frac{2\rz+\rinf}{3}\right)=\frac{4(\rz+\rinf)^2}{9R}\; .
\ee
\vskip2mm
The number of examples generated by the semi--inverse method developed in the section~\ref{s6.3.1} can be easily extended. It is worthwhile to note that the Examples 1---3 can be obtained as exact solutions of the first completely integrable model of Appendix~\ref{C}, under the proper choice of the function $f(\rv)$ involved in the interaction law~(\ref{Fpropro}). However, there is a significant difference in these two approaches. Namely, the interaction law in the semi--inverse method is fixed only along the heteroclinic curve, and can be extended elsewhere without any singularities, as stated. Contrary to this, the law~(\ref{Fpropro}) is specified in the whole plane, so the singularities on the $\rv$ axis outside its $[\rinf,\rz]$ interval are present in it, making it hardly plausible form the physics standpoint. Also, because of the freedom of the interaction law extension, the first approach demonstrates how generic these non-singular solutions are.

\section{Friedmann Cosmology with Interaction between Dark Energy and Multi--Phase Matter \label{s7}}

We now consider a more realistic cosmological model with a number of different matter species (as, in particular, in the $\Lambda$CDM model); for the sake of generality, we assume the number, $N>1$, of matter phases arbitrary. Each of the phases is described by its density, $\rn=\rn(t)$, and pressure, $\pn=\pn(t)$, related by the partial equation of state:
\be\label{EOSp}
\pn=\wn\rn,\quad \wn>-1,\quad n=1,2,\ldots,N\; .
\ee
Two species with the same equation of state might still differ by other physical properties, so we {\it do not} assume $w_m\not=\wn$ for $m\not=n$.  In the presence of DE, the total density and pressure thus become:
\bea\label{totNph}
\rt=\rv+\rho=\rv+\sum\limits_{n=1}^{N}\,\rn\; ;\nonumber\qquad\qquad\\
\pt=\pvc+p=\pvc+\sum\limits_{n=1}^{N}\,\pn=-\rv+\sum\limits_{n=1}^{N}\,\wn\rn\,.
\eea

Cosmological evolution is again described by the Friedmann equations~(\ref{FrEq}) but with the total density and pressure~(\ref{totNph}). The first Friedmann equation defines the time dependence of the scale factor (or the co-moving volume $V(t)=a^3(t)$) by the formula~(\ref{volt}). The second one,  the equation of energy conservation in the form~(\ref{FrEqV}), $d\rt/dV=-(\rt+\pt)/V$, turns to
\[
\frac{d}{dV}\left(\rv+\sum\limits_{n=1}^{N}\,\rn \right) =  -\frac{1}{V}\sum\limits_{n=1}^{N}\,(1+\wn)\rn\; 
\]
(compare with the corresponding single phase equation~(\ref{key21ph})), or
\be\label{gov1Nph}
\frac{d\rv}{dV}+\sum\limits_{n=1}^{N}\,\left[\frac{d\rn}{dV} +\frac{(1+\wn)\rn}{V}\right] = 0\; .
\ee

In the usual approach each specie is assumed to be conserved,
\be\label{partcons}
\frac{d\rn}{dV} +\frac{(1+\wn)\rn}{V} = 0,\qquad n=1,2,\ldots,N\; .
\ee
Under this condition equation~(\ref{gov1Nph}) requires that the DE density is constant, and the whole cosmological solution becomes thus
\be\label{standsolN}
\rv=\mbox{const},\qquad
\rn=C_n/V^{(1+\wn)},\quad C_n>0,\quad n=1,2,\ldots,N\; .
\ee

There are enough grounds for considering matter species in our universe not interacting with each other. We retain this standard assumption, but, as everywhere in this paper, do not forbid any of them to interact with DE. In case when $N_i\geq 1$ species interact with heavy vacuum, $N_i$ equations describing this interaction should be added to the equation~(\ref{gov1Nph}), to determine  the evolution of all the relevant densities. We discuss two cases: a) when a single matter phase interacts with DE, $N_i=1$, and b) when several phases are interacting, $1< N_i\leq N$.

\subsection{Single Matter Phase Interacting with Dark Energy\label{s7.1}}

Let the matter component interacting with DE have the number $n=1$, so it is described by $\rho_1$; all other matter phases are conserved. The densities of the latter are as in equation~(\ref{standsolN}),
\be\label{standN-1}
\rn=C_n/V^{(1+\wn)},\quad C_n>0,\quad n=2,\ldots,N\; ,
\ee
and the conservation equation~(\ref{gov1Nph}) reduces to
\be\label{consinter1Nph}
\frac{d(\rv+\rho_1)}{dV}=-\frac{(1+w_1)\rho_1}{V}\; .
\ee

To find $\rho_1(t)$ and $\rv(t)$, we need an equation specifying the interaction between the two. It is natural to take it in the same general form~(\ref{key21ph}), that is,
\be\label{inter1Nph}
\frac{d\rv}{dV}= \frac{ F(\rv,\rho_1)}{V}\; ,
\ee
where $F(\rv,\rho_1)$ is some interaction function. Combining the last two equations gives the governing system
\be\label{govsinter1Nph}
\frac{d\rho_1}{dV}=  -\frac{(1+w_1)\rho_1+ F(\rv,\rho_1)}{V},\qquad\qquad
\frac{d\rv}{dV}= \frac{ F(\rv,\rho_1)}{V}\; ,
\ee
which is, naturally,  nothing else as the system~(\ref{govs1ph}) controlling the cosmology of a single matter phase  interacting with DE (up to the notations $\rho,\;w$ replaced with $\rho_1,\;w_1$). Therefore all the general features and all exact solutions found and discussed in sec.~\ref{s3}, including non-singular cosmologies of sec. ~\ref{s6.3}, remain valid for $\rho_1(t)$ and $\rv(t)$, with all other densities given by the usual expressions~(\ref{standN-1}).

Remarkably, a toy non-singular cosmological solution of sec.~\ref{s6.1}, with the initial jump in the DE density (see fig.~1) and the linear interaction law~(\ref{GenLinF}), 
\[
 F(\rv,\rho_1)=-s\rho_1+\theta(\rv-\rinf)\; ,
\]
acquires a physical meaning due to the presence of other matter species. Namely, the initial jump in the DE density from $\rinf$ to $\rz$ can be explained by a phase transition between the matter phases~(\ref{standN-1}) (otherwise not interacting with DE) and the heavy vacuum that keeps the total energy conserved. The corresponding solution not violating, unlike the solution~(\ref{Mod3LinPhys}), energy conservation, is:
 \bea\label{InitJump}
 \rv=\rinf,\;\; \rho_1=0,\;\;\rn=C_n/V^{(1+\wn)},\; C_n>0,\; n=2,\ldots,N,\quad\mbox{for}\; 0<V<\Vst\;; \nonumber\\
\rv=\rho_\infty +{|s|}\rst\left[Q_2\left(\frac{\Vst}{V}\right)^{|\mu_{2}|}+Q_1\left(\frac{\Vst}{V}\right)^{|\mu_{1}|}\right],\;\rho_1=\rst\left[\left(\frac{\Vst}{V}\right)^{|\mu_{2}|}-\left(\frac{\Vst}{V}\right)^{|\mu_{1}|}\right]\; ,\nonumber\\
\rn=C_n^{'}/V^{(1+\wn)},\; C_n^{'}>0,\; n=2,\ldots,N,\quad \mbox{for}\quad\Vst<V<+\infty\; ;\qquad\qquad\\
\Delta\rt\Bigl|_{V=\Vst}=\left(\Delta\rv+\Delta\rho\right)\Bigl|_{V=\Vst}=
\rz-\rinf+\sum\limits_{n=2}^N\, \frac{C_n^{'}- C_n}{\Vst^{1+\wn}}=0\nonumber\; .
\eea
The constants $\mu_{1,2}$ and $Q_{1,2}$ are defined in the equations~(\ref{RootsLin}) and~(\ref{Mod3LinSol}), respectively, the interaction parameter $s$ is in the physical range~(\ref{srange}), and $\rz=\rv({\Vst+0})$. Note that some of the matter phases can turn to DE ($C_n^{'}<C_n$) at the jump,  some might gain  from DE ($C_n^{'}>C_n$), and some might stay unchanged ($C_n^{'}=C_n$), provided that the last equality holds. That is, since $\Delta\rv=\rz-\rinf$ is positive, $\Delta\rho=-\Delta\rv$ must be negative. Note also that the (partial) phase trajectory of this solution in the plane $\{\rv,\;\rho_1\}$ remains as shown in~fig.~1. 

The universe~(\ref{InitJump}) starts ($t,V\to+0$) with $\rho_1=0$, $\rv=\rinf$, and all other matter phases  singular, undergoes an instant matter--DE phase transition raising $\rv$ to $\rz$ at some moment $t=t_*\;(\Vst=V(t_*))$, and drives finally  ($t,V\to\infty$) to the initial de Sitter universe with $\rv=\rinf$ and $\rho=0$.

\subsection{Any Number of Matter Phases Interacting with Dark Energy\label{s7.2}}

Let now more than one matter phases interact with DE, so that the number of the interacting species is $N_i,\;1<N_i\leq N$. The densities of all other, non-interacting, matter species are again given by the standard expressions
\be\label{standN1}
\rn=C_n/V^{(1+\wn)},\quad C_n>0,\quad n=N_i+1,\ldots,N\; ,
\ee
and $N$ should be replaced with $N_i$ in the energy conservation equation~(\ref{gov1Nph}). Keeping the assumption that the matter species do not interact with each other, we take the following law of their interaction with DE ($\Fn$ is an arbitrary function):
\[
\frac{d\rn}{dV}+\frac{(1+\wn)\rn}{V}= -\frac{ \Fn(\rv,\rn)}{V},\qquad n=1,2,\ldots N_i\; ;
\]
each phase is conserved if and only if $\Fn=0$. With this, the  energy conservation equation turns to
\[
\frac{d\rv}{dV}= \frac{ F(\rv,\rho_1,\rho_2,\ldots,\rho_{N_i})}{V}\; ,
\]
where
\be\label{F}
 F(\rv,\rho_1,\rho_2,\ldots,\rho_{N_i})=\sum\limits_{n=1}^{N_i}\,\Fn(\rv,\rn)\; .
\ee
The last two equations govern the evolution of the universe in this case; as before, it is convenient to use them in an autonomous form, 
\bea\label{govsinterN1Nph}
\frac{d\rn}{d\lambda}=  -\left[(1+\wn)\rn+ \Fn(\rv,\rn)\right],\qquad n=1,2,\ldots N_i\; ; \nonumber\\
\frac{d\rv}{d\lambda}= F(\rv,\rho_1,\rho_2,\ldots,\rho_{N_i}):\qquad\lambda=\ln\left(V/\Vst\right)\; ,\qquad\quad
\eea
with $F$ defined by the equality~(\ref{F}). 

Generally, this is a nonlinear autonomous system of ODEs of the order\break $N_i+1\geq3$, which allows for  solutions with various behavior: even a strange attractor is possible, in principle, in the large time limit. This alone shows that the approach in which matter is represented by a single `dominant' component (like radiation, $w=1/3$, in our early universe, or dark matter, $w=0$, later) might be insufficient no matter how small the abundances of other matter species are. 

A usual regular limiting behavior at large times occurs when a cosmological solution goes to a rest point $P=\{\rho_1^{*},\rho_2^{*},\ldots,\rho_{N}^{*},\rvst\}\equiv P\{\rnst,\rvst\}$ of the system~(\ref{govsinterN1Nph}). Such a rest point is described by the equations:
\[
\rnst=(1+w_1)^{-1}\Fn(\rvst,\rnst),\quad n=1,2,\ldots N_i; \quad \sum\limits_{n=1}^{N_i}\,\Fn(\rvst,\rnst)=0.
\]
It is a physical equilibrium when  $\rnst\geq0$, therefore in this case $\Fn(\rvst,\rnst)\geq0$ for all relevant values of $n$. But then the second of the above equations implies then $\Fn(\rvst,\rnst)=0$ for all $n$, so from the first equation it follows that the only possible physical rest point is
\be\label{restN1Nph}
\rnst=0, \qquad \Fn(\rvst,0)=0, \qquad n=1,2,\ldots N_i; \qquad \rvst>0\; ;
\ee
it corresponds again to a de Sitter universe. If an equilibrium point exists and is stable, then, in view of the expressions~(\ref{standN1}),  a set of cosmological solutions of a non--zero measure tends to it at large times. However, this requires all interaction functions $\Fn(\rv,0),\;n=1,2,\ldots N_i,$ have a common positive root $\rvst$.

This is a strong restriction, unless some serious physics underlies it; if it is not valid, then  the densities of the interacting matter species do not all tend to zero at large times, and the DE density does not tend to a constant. So generically the large time behavior of cosmological solutions with several matter phases involved in the DE--matter  interaction is more complicated than the usual one; this is a characteristic feature of the multiple phase interaction.

Of course, the governing system~(\ref{govsinterN1Nph}) cannot be explicitly integrated for a general set of interaction laws $\Fn$. For this reason, below we explore two more particular models of interaction allowing for a detailed analysis and some new features.

\subsubsection{Linear Interaction Laws\label{s7.2.1}}

In a complete similarity with the case of single matter phase (see formula~(\ref{GenLinF})) we consider linear interaction laws 
\be\label{linNph}
\Fn(\rv,\rn)=-\sn\rn+\then(\rv-\rinf),\quad \sn,\then,\rinf=\mbox{const},\quad\rinf\geq 0\; .
\ee
The governing equations~(\ref{govsinterN1Nph}) become thus
\bea\label{}
\frac{d\rn}{d\lambda}=  -\left[(1+\wn-\sn)\rn+\then(\rv-\rinf)\right],\qquad n=1,2,\ldots N_i\; ; \nonumber\\
\frac{d\rv}{d\lambda}= -\sum\limits_{n=1}^{N_i}\sn\rn+\Theta(\rv-\rinf),\quad \Theta=\sum\limits_{n=1}^{N_i}\then\; .\nonumber
\qquad\qquad
\eea
Introducing an $N_i+1$--dimensional vector function $\bz(t)$,
\be\label{bzt}
\bz(\lambda)=\{\rho_1(\lambda),\;\rho_2(\lambda),\;\ldots,\;\rho_{N_i}(\lambda),\;[\rv(\lambda)-\rinf]\}^T\; ,
\ee
we rewrite this system of the first order equations in a matrix form ($\delta_{jk}$ is the Kronecker symbol):
\bea\label{dotbzt}
\frac{d\bz}{d\lambda}={\cal M}\bz\; ;\quad\qquad\qquad\qquad\qquad\qquad\\
{\cal M}_{nj}=-(1+\wn-\sn)\delta_{nj}-\then\delta_{jN_{1}+1},\;n=1,2,\ldots N_i\;;\nonumber\\
{\cal M}_{N_{1}+1j}=-s_j,\;n=1,2,\ldots N_i;\quad {\cal M}_{N_{1}+1N_{1}+1}=\Theta\; .\quad\nonumber
\eea
It is a linear system with constant coefficients, so its general solution is obtained as a linear combination of exponents of $\lambda$ (powers of $V$):
\be\label{bzsol}
\bz(\lambda)=\sum\limits_{k=1}^{N_i+1}A_k\bbe_k\exp(\mu_k\lambda)=
\sum\limits_{k=1}^{N_i+1}B_k\bbe_k V;\qquad B_k=A_k/\Vst^{\mu_k}\; .
\ee
Here $\mu_k$ are the eigenvalues of the matrix ${\cal M}$, that is, the roots of the algebraic equation
\[
\mbox{det}\left({\cal M}-\mu I\right)=0\quad(I\;\mbox{is the unit matrix})\; ,
\]
and ${\bbe_k}$ are the corresponding normalized eigenvectors,
\[
\left({\cal M}-\mu I\right)\bbe_k=0,\;\;\bbe_k=\{e_{k1},\;\ldots,\;e_{kN_i+1}\}^T,\;\;\sum\limits_{n=1}^{N_i+1}e_{kn}^2=1,\;\; k=1,\ldots N_i+1\;
\]
(for brevity, we consider only the generic case when all $\mu_k$ are different). 

The matrix ${\cal M}$ is not symmetric, so its eigenvalues might be complex, coming in complex conjugate pairs. Since the physical solution must be real, real parts should be taken at the proper places of expression~(\ref{bzsol}). Namely, suppose there are $N_c\geq1$ pairs of complex eigenvalues $\mu_k$ and $\bar\mu_k$, with the eigenvectors $\bbe_k$ and $\bar\bbe_k$, respectively, $k=1,2,\ldots,N_c$. The physical solution then becomes
\bea\label{bzsolphys}
\bz(\lambda)=\sum\limits_{k=1}^{N_c}C_kV^{\eta_k}\left[\br_k \cos(\nu_k\ln V)+
\bbi_k\sin(\nu_k\ln V)\right]+\sum\limits_{k=2N_c+1}^{N_i+1}C_kV^{\mu_k}\bbe_k;\nonumber\\
 C_k=A_k/\Vst^{\eta_k};\qquad\eta_k=\mbox {{\bf Re}}(\mu_k),\quad \nu_k=\mbox {{\bf Im}}(\mu_k)\;;\qquad\qquad\quad\\
  \br_k=2{\mbox{{\bf Re}}}(\bbe_k)/\Vst^{\eta_k},\,\bbi_k=-2{\bf\rm Im}(\bbe_k)/\Vst^{\eta_k}\;, \qquad\qquad\qquad\;\;\nonumber
\eea
but it still requires two additional conditions to be met. First, all the densities must vanish at large times $(V\to\infty)$, so all the powers of $V$ must be negative,
\be
\eta_k=\mbox {{\bf Re}}(\mu_k)<0,\;\;1\leq k\leq N_c; \qquad \mu_k<0,\;\;2N_c+1\leq k\leq N_i+1\; ;
\ee
this shows also that the expansion starts $(V\to+0)$ from singularity. 

The second and more constraining condition comes from the fact that all the densities must be positive throughout the expansion. For the case of a single matter specie interacting with DE, corresponding to $N_i=2$, this condition never holds, as demonstrated in section~\ref{s4}. For $N_i>2$ this condition might be possible to meet with some proper combination of parameters $\wn,\;\sn,\;\then$, and the right choice of the arbitrary constants $C_k$. Additional restrictions are needed when some of the eigenvalues $\mu_k$ are indeed complex ($N_c\geq1$). In this case the densities contain some terms oscillating around zero (the first sum in the expression~(\ref{bzsolphys})); those oscillations must be dominated by other strictly positive monotonic contributions. This can happen if one of the real negative eigenvalues $\mu_k$ is smaller than all $\eta_k,\;k=1,2,\ldots,N_c$, and the other one is larger than them. If this is true, the oscillations are compensated at least near the initial singularity $(V\to+0)$ and towards the end of the expansion $(V\to\infty)$, with a possibility for the density to stay positive in between as well.

If all the mentioned conditions are fulfilled, then the physical solution is
\bea\label{densphys}
\rho_n=\sum\limits_{k=1}^{N_c}C_kV^{-|\eta_k|}\left[r_{kn} \cos(\nu_k\ln V)+
i_{kn}\sin(\nu_k\ln V)\right]+\sum\limits_{k=2N_c+1}^{N_i+1}C_k e_{kn}V^{-|\mu_k|};\nonumber\\
\rv=\rinf+\sum\limits_{k=1}^{N_c}C_kV^{-|\eta_k|}\left[r_{kN_i+1} \cos(\nu_k\ln V)+
i_{kN_i+1}\sin(\nu_k\ln V)\right]+\qquad\\
+\sum\limits_{k=2N_c+1}^{N_i+1}C_k e_{kN_i+1}V^{-|\mu_k|}\; .\qquad\qquad\qquad\qquad\qquad\nonumber
 \eea
This solution drives to a de Sitter universe $\rn=0,\;\rv=\rinf$. If oscillations are present, then their frequency becomes infinitely large both at the initial singularity and the expansion end.

It is worthy to consider one particular case studied in detail in section~\ref{s4} for the single matter phase cosmology.  In this case the linear interaction law does not depend on the DE density, i.e., $\then=0$, $\Fn(\rv,\rn)=-\sn\rn$. Thus every density $\rn$ satisfies its own linear equation, making the answer rather simple:
\be\label{rhononly}
\rho_n=\frac{C_n}{V^{1+\wn-\sn}},\;\;n=1,2,\ldots N_i;\qquad
\rv=\rinf+\sum\limits_{k=1}^{N_i}\,\frac{\sn C_n}{V^{1+\wn-\sn}}\; ;
 \ee
it is an exact analog of the single matter phase solution~(\ref{Mod1sol}), with all the properties described in section~\ref{s4}. The solution~(\ref{rhononly}) is physically meaningful under the condition
\[
0<\sn<1+\wn,\qquad n=1,2,\ldots N_i \; ,
\]
which is a generalization of the condition~(\ref{bzsolphys}). It guarantees that all the densities, including $\rn$, are positive and monotonically decreasing with matter vanishing at infinity. However, the left inequality above is, in fact, necessary for only one value of $n$, say, $n=k$, corresponding to the maximum difference $(\wn-\sn)$,
\[
w_k-s_k=\max_{1\leq n\leq N_i}(\wn-\sn)\; ,
\]
providing that the DE density is positive at small times, near the singularity. Depending on the values of the positive constants $C_n$, some of other parameters $\sn,\;n\not=k$, can be negative, with $\rv$ remaining positive throughout the expansion. In this case it might be non-monotonic, having positive maxima and minima at some moments of time.

\subsubsection{Quadratic Interaction Laws\label{s7.2.2}}

Here we set
\be\label{qNph}
\Fn(\rv,\rn)=-(\sn/R)\rn(\rv-\rz),\quad \sn, R,\rz=\mbox{const},\quad R,\rz\geq 0\; .
\ee
The governing equations~(\ref{govsinterN1Nph}) become:
\bea\label{qgovNph}
\frac{d\rn}{d\lambda}+(1+\wn)\rn=\frac{\sn}{R}\rn(\rv-\rz),\qquad n=1,2,\ldots N_i\; ; \nonumber\\
\frac{d\rv}{d\lambda}= -\frac{1}{R}\,\sum\limits_{n=1}^{N_i}\sn\rn(\rv-\rz)\; .\qquad
\qquad\qquad
\eea
As usual, we are interested only in its physical solutions, with all the densities positive and matter phase densities vanishing at the end of the expansion.

The autonomous system~(\ref{qgovNph}) of $(N_i+1)$ equations has a physical rest point $\rn=0,\rv=\rinf$ with any $\rinf\geq0$, i.e., the whole semi-axis $\rv\geq0$ consists of its equilibriia.  They can attract solutions at large times; the corresponding asymptotic expressions for the case $\rinf\not=\rz$ are ($t,V\to+\infty$):
\bea\label{qLargetime}
\rn=\frac{D_n}{V^{1+\wn+\gamma\sn}}\left[1+o(1)\right],\;\;\gamma=\frac{\rz-\rinf}{R},\;\;\rinf\not=\rz;\;\;
n=1,2,\ldots N_i \; ; \qquad\\
\rv=\rinf-\frac{\gamma s_k D_k}{(1+w_k+\gamma s_k) V^{1+w_k+\gamma s_k}}\left[1+o(1)\right],\;\;
w_k+\gamma s_k=\min_{1\leq n\leq N_i}(\wn+\gamma\sn)\; ;\nonumber
\eea
here $D_n>0$ is some constant. For the matter densities to vanish asymptotically, the following condition is required, for all $n$:
\[
1+\wn+\gamma\sn>0\; ,
\]
which splits into two sets of inequalities:
\be\label{qdensvanish}
a)\,\rinf>\rz,\;\sn<\frac{R}{\rinf-\rz}(1+\wn);\quad b)\,\rinf<\rz,\;\sn>-\frac{R}{\rz-\rinf}(1+\wn)\; .
\ee
In both cases the signs of the parameters   $\sn$ are not fixed: some of them can be positive, the other can be negative.  

The large time asymptotics for the exceptional case of the attracting rest point with  $\rv=\rz$ is more complicated, except  the obvious exact solution with the constant DE density:
\be\label{qrv=rz}
\rn={D_n}/{V^{1+\wn}},\;n=1,2,\ldots N_i ;\qquad  \rv=\rz=\mbox{const}\; .
\ee
Here there is no interaction between dark energy and matter, thus all the matter species are conserved (by the formulas~(\ref{qrv=rz}) and~(\ref{standN1})). However, this solution, and thus the parameter $\rz$, can play a role in the initial behavior of solutions that may `branch' from the above one out of the singularity. The asymptotic formulas describing such behavior are ($t,V\to+0$):
\bea\label{qsmalltime}
\rn=\frac{D_n}{V^{1+\wn}}\left[1+o(1)\right],\quad D-n>0,\quad n=1,2,\ldots N_i \; ; \qquad\qquad\\
\rv=\rz+D_0\exp\left[\frac{-s_kD_k}{(1+w_k)V^{1+w_k}}\right]\left[1+o(1)\right],\;\;
w_k=\max_{1\leq n\leq N_i}\wn,\;\; s_k>0\; .\nonumber
\eea
The last inequality is needed because the correction to $\rz$ must vanish in the limit. For the first time this correction proves to be exponentially small; all other cosmological solutions obtained and discussed so far do not have this feature.

A solution with the small time asymptotics~(\ref{qsmalltime}) is similar to the solution~(\ref{qrv=rz}) in a sense that all matter in the universe described by it is born form a singularity, and the DE density is finite at the initial moment of time. If this solution also has the large time behavior described by the formulas~(\ref{qLargetime}), then the DE density evolves from one value, $\rz$, in the beginning, to some other, $\rinf$, at the end of the expansion.

The governing system~(\ref{qgovNph}) of $(N_i+1)$ equations can be reduced to just two equations for any $N_i>1$, since its  $N_i-1$ integrals are explicitly found. Indeed, the first equation~(\ref{qgovNph}) implies
\[
\frac{d\ln\rn}{d\lambda}+(1+\wn)=\frac{\sn}{R}(\rv-\rz),\qquad n=1,2,\ldots N_i\; ,
\]
 allowing for the following $N_i-1$ combinations:
\[
s_n\frac{d\ln\rho_1}{d\lambda}-s_1\frac{d\ln\rho_n}{d\lambda}+\left[s_n(1+w_1)-s_1(1+w_n)\right]=0,
\quad n=2,3,\ldots N_i\;. 
\]
This equations can be immediately integrated to give the expressions for all interacting phase densities through the first one,
\bea\label{intqNph}
\rn=A_n\rho_1^{\then}\exp(\beta_n\lambda)=B_n\rho_1^{\then}V^{\beta_n},\quad B_n=A_n/\Vst^{\beta_n},\quad  n=2,3,\ldots N_i\; ;\nonumber\\
 A_n,B_n>0;\qquad\then=s_n/s_1,\qquad\beta_n=-(1+w_n)+(s_n/s_1)(1+w_1)\;.\;
\eea
What remains is the system~(\ref{qgovNph}) of two equations for $\rho_1$ and $\rv$, the second of them with the coefficient depending general on the evolution variable ($\lambda$ or $V$):
\bea\label{qgov2eqNph}
\frac{d\rho_1}{d\lambda}+(1+w_1)\rho_1=\frac{s_1}{R}\rho_1(\rv-\rz),\quad 
\frac{d\rv}{d\lambda}= -\frac{ {\cal P}(\rho_1,\lambda)}{R}\,(\rv-\rz)\; ; \qquad\\
{\cal P}(\rho_1,\lambda)=\sum\limits_{n=1}^{N_i}\sn A_n\rho_1^{\then}\exp(\beta_n\lambda)
=\sum\limits_{n=1}^{N_i}\sn B_n\rho_1^{\then}V^{\beta_n}\; .\qquad\qquad\qquad\nonumber
\eea
Here by the definition~(\ref{intqNph}) $\beta_1=0$, $\theta_1=s_1/s_1=1$, and we set $A_1=1$; all the densities $\rho_2,\rho_3,\ldots,\rho_{N_i}$ are replaced with their expressions~(\ref{Q}). 

A physically meaningful solution of the system~(\ref{qgov2eqNph}) together with the expressions~(\ref{intqNph}) provides the complete answer, i.e., a cosmological solution describing the universe with $N$ matter species, of which $N_i>1$ interact with dark energy by the law~(\ref{qNph}).

We consider now one example where the equations integrate completely, that is, the second order system~(\ref{qgov2eqNph}) proves to be explicitly integrable. This is the case when the equations~(\ref{qgov2eqNph}) become autonomous, i.e., the independent variable is not involved in the second of them. We take (see formulas~(\ref{qgov2eqNph}))
\[
\beta_n=0,\qquad \sn=s_1\,\frac{1+\wn}{1+w_1},\qquad n=2,3,\ldots N_i\; ;
\]
note that all parameters $\sn$ are of the same sign. Using this in the definition~(\ref{qgov2eqNph}) of the function ${\cal P}$ we find
\be\label{PtoQ}
{\cal P}(\rho_1,\lambda)=s_1\,\sum\limits_{n=1}^{N_i}\frac{1+\wn}{1+w_1}A_n\rho_1^{\frac{1+\wn}{1+w_1}}\equiv s_1Q(\rho_1)\; ,
\ee
so the equations~(\ref{qgov2eqNph}) become:
\bea\label{qgovex}
\frac{d\rho_1}{d\lambda}+(1+w_1)\rho_1=\frac{s_1}{R}\rho_1(\rv-\rz),\quad 
\frac{d\rv}{d\lambda}= -\frac{ s_1}{R}\,Q(\rho_1)(\rv-\rz)\; .
\eea
Dividing the first of them by the second one we arrive to the equation with the separable variables,
\[
\frac{d\rho_1}{d\rv}=\frac{\rho_1}{Q(\rho_1)}\left[\frac{(1+w_1)R}{s_1(\rv-\rz)}-1\right]\; ,
\]
whose integral, by virtue of the expression~(\ref{PtoQ}), is: 
\[
s_1\,\sum\limits_{n=1}^{N_i}A_n\rho_1^{\frac{1+\wn}{1+w_1}}=K-\rv+\frac{(1+w_1)R}{s_1}\ln|\rv-\rz|\; ;
\]
here $K$ is a constant of integration. It is determined from the large time behavior of a physical solution: in this limit $\rho_1$ must vanish, and $\rv$ must tend to some value $\rinf\geq0,\,\rinf\not=\rz$. The l.h.s of the above equality goes to zero in this limit, therefore the same must happen with the r.h.s, which gives
\[
K=\rinf-\frac{(1+w_1)R}{s_1}\ln|\rinf-\rz|\; ,
\]
and the integral becomes
\be\label{qint}
\sum\limits_{n=1}^{N_i}A_n\rho_1^{\frac{1+\wn}{1+w_1}}=\rinf-\rv+
\frac{(1+w_1)R}{s_1}\ln\left|\frac{\rv-\rz}{\rinf-\rz}\right|\; .
\ee
It is convenient to treat this as an equation for the DE density as function of the matter density depending also on the limit value $\rinf$. If it has a solution $\rv=\rv(\rho_1,\rinf)$ such that $\rv(0,\rinf)=\rinf$, then the first equation equation~(\ref{qgovex}) reduces to integrating a known function whose result determines $\rho_1(\lambda)$, or $\rho_1(V)$:
\be\label{qint2}
\int\limits^{\rho_1}\frac{dx}{x\left\{1+w_1+\left(s_1/R\right)\left[\rz-\rv(x)\right]\right\}}=-\lambda=\ln\left(\frac{\Vst}{V}\right)\; .
\ee
If, in its turn, this equation has a solution $\rho_1=\rho_1(V)$ going to zero when $V\to+\infty$, then we have a consistent solution to the system~(\ref{qgovex}). This is a cosmological  solution if both densities are positive on the whole semi--axis $V>0$.

A simple enough graphic analysis of the transcendental equation~(\ref{qint}) shows that its positive solution $\rv=\rv(\rho_1,\rinf)$ does exist under certain restriction on the parameter values. First of all, the large time limit  of DE density should be smaller than $\rz$,
\be\label{rinfrz}
\rinf<\rz\; .
\ee
Under this condition there are two cases yielding solutions of a different type.
\vskip2mm
\underline{Case A}  Parameter $s_1$, and hence all $\sn$, are positive, 
\be\label{s1pos}
\sn>0,\qquad n=1,2,\ldots,N_i\; .
\ee
A single solution $\rv$ to the equation~(\ref{qint}) then exists that  increases from the initial zero value to $\rinf$ in the course of the expansion, while the density $\rho_1$ decreases from a  finite initial value $\rst$ to zero (the value $\rst$ is found from the equation~(\ref{qint}) with $\rv=0$).This is not surprising, because the governing equations~(\ref{qgovNph})  show that in this case the interaction reduces the matter phases and produces heavy vacuum for $0<\rv<\rinf<\rz$.  According to the expressions~(\ref{intqNph}), all other interacting matter densities $\rn,\, n=2,3,\ldots,N_i$ are also finite at the beginning of the expansion.

However, equation~(\ref{qint2}) shows that the finite initial value of  $\rho_1$ corresponds to a finite non-zero initial value of $V$, or of the scale factor, which does not make sense, unlike the situation described at the end of section~\ref{s6.2}. So we need to extend the solution towards larger density $\rho_1$ (smaller values of $V$), but the DE density becomes negative there, for $\rho_1>\rst$. The solution thus has no physical meaning.
\vskip2mm
\underline{Case B}  Parameter $s_1$, and hence all $\sn$, are negative, 
\be\label{s1neg}
\sn<0,\qquad n=1,2,\ldots,N_i\; .
\ee
Here the positive solution $\rv$ to the equation~(\ref{qint}) decreases from the initial value to $\rz$ to $\rinf$, while the matter density $\rho_1$, singular at the beginning, decreases monotonically to zero. Dark energy permanently produces the interacting matter phases while being reduced accordingly, which production  slows down the decay of matter densities, as compared to the absence of the interaction. 

By the expressions~(\ref{intqNph}), all interacting densities $\rn,\, n=1,2,\ldots,N_i,$ also emerge from the initial singularity; since the DE density is finite, this is a cosmology of a mixed, singular---non-singular, type. From the equation~(\ref{qint2}) we find that the initial behavior of the solution is given by the formulas~(\ref{qsmalltime}), and its final behavior is described by the expressions~(\ref{qLargetime}). So each matter density is inversely proportional to some power of $V$, or the scale factor, at the beginning of the expansion, and to some {\it other} power at its end. The initial dependencies are the same as in the case without the interaction, because it becomes negligibly small when $\rv\to\rz+0$.

Finally, we note briefly the general quadratic interaction law
\be\label{qGenNph}
\Fn(\rv,\rn)=a_n\rn^2+b_n\rv^2+c_n\rn\rv+d_n\rn+e_n\rv\; ,
\ee
with some constants $a_n,\;b_n,\;c_n,\;d_n$, and $e_n$. The condition for a physical equilibrium point is
\[
b_n\rst^2+e_n\rst=0,\quad \rst\geq0,\quad, n=1,2,\ldots,N_i\; .
\]
So an empty space is always a rest point, but the existence of a non-trivial de~Sitter equilibrium requires
\[
\rst=-b_n/e_n>0,\qquad n=1,2,\ldots,N_i\; ,
\]
giving $N_i$ relations on the $5N_i$ parameters involved. For small values of matter densities, i.e., in the large time limit, the solution is effectively governed by the general linear law. Otherwise the signature of the quadratic form in the r.h.s. of equation~(\ref{intqNph}) is most important for the existence of physical solutions and their properties.

\subsubsection{Non-Singular Cosmologies\label{s7.2.3}}

Non-singular cosmologies found in section~\ref{s6} for one matter specie exist in the multiple matter component case as well. They evolve according to the general picture of non-linear interaction  described there, namely, as heteroclinic phase trajectories connecting one physical rest point, $\rn=0,\;\rv=\rz>0$, of the system~(\ref{govsinterN1Nph})  with the other, $\rn=0,\;\rv=\rinf>0$, now in the $N_i+1$--dimensional phase space. 

Note that if not all matter species interact with dark energy ($N_i<N$), then a `mixed' type cosmology is obtained in this way: the interacting components and DE are non-singular, but the non-interacting ones start at a singularity. In this case there is no limitations on the spacetime curvature pointed out in section~\ref{s6.2}, because the denisities of the conserved components dominate everything else, including the curvature contribution, at the expansion beginning. The corresponding `mixed type' universe can be either open, or flat, or closed. When all the matter species are interacting, $N_i=N$, then an entirely  non-singular universe is necessarily open.

The semi--inverse method for constructing such solutions developed in the section~\ref{s6.3.1} also works in the general case. Indeed, in a complete similarity with the one-specie {\it Anzatz}~(\ref{rhoh}) we assume that a heteroclinic trajectory ${\cal H}$ is described by the equations
\be\label{calH}
\rn=\hn(\rv),\;\;\rinf<\rv<\rz,\;\; \hn(\rz)=\hn(\rinf)=0;\quad n=1,2,\ldots N_i\; ,
\ee
where the functions $\hn$, positive inside their domain, are otherwise arbitrary. The appropriate calculations go the same way as in the section~\ref{s6.3.1}. 

Namely, the first $N_i$ equations~(\ref{govsinterN1Nph}) require certain values of the interaction functions $\Fn(\rv,\rn)$ on the heteroclinic curve which are found from the linear algebraic system (as usual, the prime denotes the derivative in $\rv$):
\bea
-(1+w_k)h_k(\rv)=F_k(\rv,h_k(\rv))+h_k^{'}(\rv)\sum\limits_{n=1}^{N_i}\,\Fn(\rv,\hn(\rv))\;,\nonumber\\
\rinf<\rv<\rz,\qquad k=1,2,\ldots N_i\; .\qquad\qquad\qquad\nonumber
\eea
It allows for a simple explicit solution: by summing up all the equations, we first find the sum
\be\label{S}
S(\rv)\equiv F\biggl|_{\cal H}=\sum\limits_{n=1}^{N_i}\,\Fn\biggl|_{\cal H}=
-\frac{\sum\limits_{n=1}^{N_i}\,(1+\wn)\hn(\rv)}{1+\sum\limits_{n=1}^{N_i}\,\hn^{'}(\rv)}\;,
\ee
and then, from each of the above equations, functions $\Fn$ on the curve ${\cal H}$:
\[
F_k\biggl|_{\cal H}=--(1+w_k)h_k(\rv)-S(\rv),\qquad k=1,2,\ldots N_i\; 
\]
(we do not actually use them in what follows). They have no singularity on the interval $[\rinf,\;\rv]$ under the condition
\be\label{nosingNph}
\min_{\rinf\leq\rv\leq\rz}\sum\limits_{n=1}^{N_i}\,\hn^{'}(\rv)>-1\; ,
\ee
and can be extended from the curve ${\cal H}$ to the whole phase space in a continuum of ways, as noted in section~\ref{s6.3.1}; the inequality~(\ref{nosingNph}) is a direct generalization of the single-phase condition~(\ref{nosing}). 

Now, by the formula~(\ref{S}), the last of the governing equations~(\ref{govsinterN1Nph}) on the heteroclinic curve ${\cal H}$ becomes
\[
\frac{d\rv}{d\lambda}\biggl|_{\cal H}=F\biggl|_{\cal H}=S(\rv)=-\frac{\sum\limits_{n=1}^{N_i}\,(1+\wn)\hn(\rv)}{1+\sum\limits_{n=1}^{N_i}\,\hn^{'}(\rv)}\; ,
\]
so determining $\rv$ reduces to integrating the known function. The result, in terms of the variable $V$, is:
\be\label{rvNph}
\exp\left[\hat H(\rv)\right]=\frac{C}{V},\qquad \hat H(\rv)=
{\Large\int\limits^{\rv}}\,\frac{1+\sum\limits_{n=1}^{N_i}\,\hn^{'}(x)}{\sum\limits_{n=1}^{N_i}\,(1+\wn)\hn(x)}\, dx\; , 
\ee
with $C>0$ being a constant of integration. This is the analog of the equation~(\ref{rvac}) for determining the DE density. If this transcendental equation has a positive solution $\rv=\rv(V)$ decreasing monotonically from $\rv=\rz$ to $\rv=\rinf$, then $\rn=\hn(\rv(V))$, and these $N_i+1$ functions provide a solution of the system~(\ref{govsinterN1Nph}) corresponding to the heteroclinic trajectory ${\cal H}$ in its phase space. If exist, the densities of non-interacting species are given by the usual expressions~(\ref{standN1}), completing the solution describing a `mixed' cosmology.

By specifying algebraic behavior of functions $\hn$ at the ends of the interval of their definition, like in the equalities~(\ref{asymph}), one can find the asymptotic behavior of the interacting densities at the beginning and end of the expansion, first as functions of $V$, as it is done in the Appendix~\ref{A}, and then as functions of time, as in the formulas~(\ref{ast0}),~(\ref{astinf}).

We here extend our calculations for just one special case,  which leads to even more similarity with the results of section~\ref{s6.3.1}, and hence to the set of particular exact solutions.  Namely, we assume that the projections of the heteroclinic trajectory ${\cal H}$ on each of the planes $\{\rn,\rv\},\; n=1,2,\ldots,N_i$ all have the same shape. That is, we assume that $\hn(\rv)$ differ from each other only by scaling:
\bea\label{samehn}
\hn(\rv)=\chi_n h(\rv), \qquad \chi_n>0\; ;\qquad\qquad\quad\\
  h(\rv)>0\;\;\mbox{for}\;\;\rinf<\rv<\rz,\qquad h(\rinf)=h(\rz)=0\; .  \nonumber
\eea
It is then straightforward to calculate,  by the formula~(\ref{rvNph}):
\[
\exp\left[\hat H(\rv)\right]=\chi^{-\frac{1}{1+\bar w}}\left\{\check h(\rv)\exp\left[H(\rv)\right]\right\},\quad
H(\rv)={\Large\int\limits^{\rv}}\,\frac{dx}{\check h(x)}\; ,
\]
where
\[
\chi=\sum\limits_{n=1}^{N_i}\,\chi_n,\quad 1+\bar w =\frac{1}{\chi}\,\sum\limits_{n=1}^{N_i}\,(1+\wn)\chi_n,\quad
\check h(\rv)=\chi  h(\rv)\; .
\]
After some constant reassignment we can thus rewrite the resolving equation~(\ref{rvNph}) in exactly the form of the resolving equation~(\ref{rvac}) of the single--phase case:
\be\label{rvacNphsamehn}
\check h(\rv)\exp H(\rv)=\rst\left(\frac{\Vst}{V}\right)^{1+\bar w}\; ,
\ee
with just $h$ sreplaced with $\check h$, and $w$ replaced with $\bar w$. So one can use the exact solutions of the examples from section~\ref{s6.3.2} obtained for the fucntions
\bea
\check h(\rv)=(\rz-\rv)(\rv-\rinf)/R, \quad \check h(\rv)=\theta(\rz-\rv)(\rv-\rinf)/\rv\;,\nonumber\\
\check h(\rv)=(\rz-\rv)(\rv-\rinf)^2/R^2\; ,\qquad\qquad\quad\qquad\qquad\quad\nonumber
\eea
as well as construct many other.

\subsection{Three Matter Phases: a Model for Our Universe\label{s7.3}}

To get closer to the only reality known by us, we finally consider a cosmological model with dark energy and three matter phases: dark matter (DM), $w=0$, normal matter, $w=0$, and radiation, $w=1/3$. There are many speculations about a possible relation between the dark energy and dark matter, which seem plausible intuitively. Following these ideas we here assume that {\it only dark matter interacts with dark energy}, and the other two matter phases are conserved, as in the usual cosmological models. This puts us in the case of the section~\ref{s7.1} with  $N=3$; we denote $\rho_1=\rho_{dm}$ the DM density ($w_1=0$), $\rho_2=\rho_{m}$ the density of normal matter ($w_2=0$), and $\rho_3=\rho_{r}$ the density of radiation ($w_3=1/3$). The last two species are conserved, so their densities are given by the standard formulas:
\be\label{matrad}
\rho_{m}=C_m/V,\qquad \rho_{r}=C_r/V^{4/3},\qquad C_{m,r}>0 \; .
\ee
Thus there is always the Big Bang in this model, but DE and DM are not necessarily involved in it. The behavior of $\rho_1$ and $\rv$ is determined by the system~(\ref{govsinter1Nph}), written as
\[
\frac{d\rho_1}{dV}=  -\frac{\rho_1+ F(\rv,\rho_1)}{V},\qquad\qquad
\frac{d\rv}{dV}= \frac{ F(\rv,\rho_1)}{V}\; ,
\]
or, in terms of $\lambda=\ln(V/\Vst)$ and $\rho_{dm}=\rho_1$, as
\be\label{govsinter13ph}
\frac{d\rho_{dm}}{d\lambda}=  -\left[\rho_{dm}+ F(\rv,\rho_{dm})\right],\qquad\qquad
\frac{d\rv}{d\lambda}=F(\rv,\rho_{dm})\; .
\ee
It is nothing else as the equations~(\ref{govsaut1ph}) with $w=0$, so we can use all results of section~\ref{s3} in the discussion of our model of the Universe.

We start with a special linear interaction law~(\ref{Mod1F}), $F(\rv,\rho_{dm})=-s\rho_{dm}$, when the rate of DE reduction is proportional to the dark matter density. The corresponding exact solution~(\ref{Mod1sol}) reads:
\be\label{lin1sol}
\rho_{dm}=\frac{C_{dm}}{V^{1-s}},\qquad\rv=\rinf+\frac{s}{1-s}\,\frac{C_{dm}}{V^{1-s}}\; ;
\ee
here $C_{dm}>0,\;\rinf\geq0$ are arbitrary constants, and the interaction parameter $s$ is in the range~(\ref{srange}), $0<s<1$.

The expressions~(\ref{lin1sol}) and~(\ref{matrad}) combine to give a cosmological solution that differs from the usual one, with the constant DE density, by the power in the dependence of $\rho_{dm}$, $V^{-(1-s)}$ instead of $V^{-1}$. However, this difference is essential form the point that, although both the dark and normal matter densities tend to zero at large times, their ratio
\[
\rho_{dm}/\rho_m\propto  \left(C_{dm}/C_{m}\right)V^{s}\to\infty,\qquad V\to \infty\; ,
\]
tends to infinity at the large time limit independent of the parameters involved. So, without any fine--tuning, dark matter dominates normal matter at later stages, as observed in our universe. Otherwise,  radiation dominates the early universe, as usual,  so the scale factor $a(t)\propto t^{1/2},\; t\to+0$; non-vanishing DE dominates all other components at later time providing the typical exponential time dependence of the the scale factor (see formula~(\ref{Mod1tlarg})). 

Interestingly, this linear  model of interaction between DE and DM in our universe was checked against the observational data in a recent paper~\cite{LiZhZh}. The authors used the Planck 2013 data,
the baryon acoustic oscillations measurements, the type-Ia supernovae data, the Hubble constant
measurement, the redshift space distortions data and the galaxy weak lensing data to estimate the parameter $s$ (denoted $\beta$ in the paper). One-sigma errors of the found estimates are larger than 100\%. Generally, constraints on any interaction models are very important, but they are definitely a subject of separate paper(s).

Next, as demonstrated in section~\ref{s7.1}, the general interaction law~(\ref{GenLinF}),
\[
F(\rv,\rho_{dm})=-s\rho_{dm}+\theta(\rv-\rinf),\quad s,\theta,\rinf=\mbox{const},\quad\rinf\geq 0\; ,
\]
does not allow for any continuous physical solution. However, a solution of the form~(\ref{InitJump}) with a jump  in the DE density is possible. In this case it is given by the following expressions:
\bea\label{InitJumpOur}
 \rv=\rinf,\;\; \rho_{dm}=0,\;\;\rho_m=C_m/V,\;  \rho_{r}=C_r/V^{4/3}\quad
 \mbox{for}\; 0<V<\Vst\;;\nonumber\\
\rv=\rinf +{|s|}\rst\left[Q_2\left(\frac{\Vst}{V}\right)^{|\mu_{2}|}+Q_1\left(\frac{\Vst}{V}\right)^{|\mu_{1}|}\right],\;\rho_{dm}=\rst\left[\left(\frac{\Vst}{V}\right)^{|\mu_{2}|}-\left(\frac{\Vst}{V}\right)^{|\mu_{1}|}\right],\nonumber\\
\rho_m=C_m^{'}/V,\;\; \rho_{r}=C_r^{'}/V^{4/3}\quad \mbox{for}\quad\Vst<V<+\infty\; ;\qquad\qquad\qquad\\
\Delta\rt\Bigl|_{V=\Vst}=\left(\Delta\rv+\Delta\rho\right)\Bigl|_{V=\Vst}=
\rz-\rinf+\frac{C_m^{'}- C_m}{\Vst}+\frac{C_r^{'}- C_r}{\Vst^{4/3}}=0\nonumber\; ;
\eea
all the parameters are restricted as in the formulas~(\ref{InitJump}), and $\rz=\rv({\Vst+0})$. 

In this cosmology matter and radiation are born from a singularity on the background of a finite DE density remaining constant, $\rv=\rinf$, until some moment of time $t_*,\;\Vst=V(t_*)$. At this moment  the two existing non-interacting species undergo an instant phase transition raising the DE density to the value $\rz>\rinf$. After this the DE density relaxes all the time back  to its initial value $\rinf$, and dark matter appears whose density first grows, then reaches some maximum, and then declines to zero at infinity; the evolution of the two interacting species is depicted in fig. 1.

As before, this example leads us to non-singular cosmologies appearing under non-linear interaction laws, i.e., to the results of sections~\ref{s6.2},~\ref{s6.3}, which all apply to our current model. Non-singular cosmological solutions discussed and explicitly found there correspond to heteroclinic curves in the phase plane $\{\rv,\rho_{dm}\}$ connecting two de Sitter equlibrium states with $\rv=\rz$ and $\rv=\rinf$. So the dark energy density evolves from the initial value $\rz$ to the final value $\rinf$. Dark matter appears at the start of the evolution, its density reaches a maximum (whose value depends on the model parameters, c.f. formulas~(\ref{rmax1}),~(\ref{rmax2}),~(\ref{rmax3})) at some moment of time, and then tends back to zero. 

The conserved radiation and normal matter are born in a singularity, their densities evolve according to the usual expressions~(\ref{matrad}). At large times DM often dominates normal matter, since the former goes to zero slower than the latter. This is clearly seen from the asymptotic formulas~(\ref{asinf}) in the cases {\it a)} and {\it b)}, when at large times the ratio $\rho_{dm}/\rho_{m}$ tends to infinity independent of the model parameters. In the case {\it c)}  both densities have the same later times dependence $\propto V^{-1}$, so the DM dominance requires parameter tuning.

Of course, our universe can be also modeled with two or all the matter phases interacting with DE; the results of section~\ref{s7.2} apply to such models.

\section{Conclusion\label{s8}}

We pointed out that dark energy is not necessarily uniform if it coexists with matter: its density might vary in space and time due to the interaction between the two gravity sources.  Based on this idea, we systematically studied the Friedmann cosmology with changing cosmological constant (or DE density proportional to it), first for one matter phase (single equation of state), and then for an arbitrary number of matter species. We modeled the DE--matter interaction by specifying the rate of change of the DE density as an arbitrary function of it and the density of matter, in a single--phase case. In the case of several matter components interacting with dark energy we assumed the rate of every interacting phase density to be an arbitrary function of this density and the DE density. We thus neglected the interaction of matter phases with each other, as usual; any number of entirely non-interacting, conserved matter species might accompany the interacting ones in our model.

Within this framework we indicated some properties of cosmological solutions which hold for a general law of DE--matter interaction. We also studied numerous families of exact solutions obtained for particular interactions; some of them still contain arbitrary functions of one of the densities. 

In particular, we found singular solutions with no horizon problem in some range of parameters. Depending on the latter, the scale factor can grow in the beginning of the cosmological expansion as {\it an arbitrary large} power of time, so that one can speak about the `power inflation'. These solutions are always dominated by dark energy after some moment of time; depending on parameters, the domination might start at the singularity and continue throughout the whole expansion.

We most thoroughly investigated non-singular cosmologies (or partially non-singular, `mixed' ones, if the conserved matter components are present: their densities evolve by the usual formulas with initial singularity). We found a general mechanism of their existence. Namely, non-singular cosmological solutions are represented by heteroclinic trajectories in the phase space of interacting matter densities and the DE density. Each such trajectory connects two de Sitter universes (pure uniform dark energy) with different DE densities (a non-generic case when the initial and final density values  coincide corresponds to a homoclinic trajectory). We developed a semi--inverse method for solving the equations governing cosmological evolution that allows one to explicitly construct any number of non-singular cosmological solutions, with several examples treated in detail.

Very often different cosmological solutions exist for a given interaction law, for instance, some of them singular, and other non-singular. In a sense, this is what is called a multiverse, because those different solutions can describe many  universes existing in parallel.

We finally considered a model for our universe consisting of four components (radiation, normal matter, dark matter, and dark energy) under the assumption that only dark matter interacts with dark energy. This means that radiation and normal matter are both born in a Big Bang, while the DE and DM densities can be either singular or non-singular, in the `mixed' case. Among various properties of the considered exact solutions we note the typical domination of dark matter over the normal one at later stages of the expansion, which takes place for any values of the model parameters, without any tuning.

All these and other results were obtained strictly within the theory of general relativity, without any modifications, such as extra space--time dimensions, additional fields, etc. (recent  restrictions on such extended models derived from observations and tests are found in papers~\cite{OverdEW} - \cite{FarSal}). This fundamental physical theory remains vibrant at its centennial, despite many alternative suggestions. 

As for cosmology, the choice of the model for our universe is ultimately determined by observations. As far as our model, with dark energy and matter interacting, goes, one can think, in the very long run, about reconstructing the real  interaction law from observational data.

Our approach to general relativistic solutions with interacting dark energy and matter can be used in various other problems, starting with the classical spherically symmetric case.

\begin{acknowledgments}
I am grateful to Arthur Chernin, James Overduin, and Bob Wagoner for their valuable remarks and discussion. My special thanks go to Chernin who introduced me to cosmology more than 40 years ago, and encouraged my work on this paper.
\end{acknowledgments}
 
\appendix

\section{The $\rv(H) $ Model as a Particular Case of the Model~(\ref{key21ph})\label{O}}

Our general model~(\ref{key21ph}) of interaction between one--phase matter and DE incorporates $\rv(H) $ model of flat universe introduced in paper~\cite{ShapSol} (more references are given in section~\ref{s3}). To show this, we set the interaction function to be
\be\label{O1}
F(\rv,\rho)=F(\rv+\rho)=F(\rt)\; .
\ee
 By the first of  the Friedmann equations~(\ref{FrEq}) with $k=0$,
\be\label{O2}
\frac{3}{8\pi}H^2= \rho+\rv,\qquad H=\frac{\dot a}{a}\;,
\ee
we obtain thus
\be\label{O3}
F(\rv,\rho)=F(\rt)=\Phi(H)\;,
\ee
where $\Phi$ is arbitrary as far as $F$ is. The two equations~(\ref{govs1ph}) (energy conservation and interaction model) written in terms of $H$ are
\be\label{O4}
\dot\rho=-3H\left[(1+w)\rho+\Phi(H)\right],\qquad \dot\rv=3H\Phi(H)\;,
\ee
where the dot denotes the derivative in time, as usual; equations~(\ref{O2}),~(\ref{O4}) completely determine the cosmological expansion.

Assuming now $\rv=\rv(H)$, form the first equation~(\ref{O2}) we find the matter density as a function of $H$,
\[
\rho(H)=\frac{3}{8\pi}H^2-\rv(H)\; ,
\]
which converts the equations~(\ref{O4}) to (the prime denotes the derivative in $H$):
\be\label{O5}
\dot H=-3H\frac{(1+w)\left[\left(3H^2/8\pi\right)-\rv(H)\right]+\Phi(H)}{\left(3H/4\pi\right)-\rv^{'}(H)},\quad
\dot H=3H\frac{\Phi(H)}{\rv^{'}(H)}\; 
\ee
The compatibility condition for these two equations apparently is:
\[
-\frac{(1+w)\left[\left(3H^2/8\pi\right)-\rv(H)\right]+\Phi(H)}{\left(3H/4\pi\right)-\rv^{'}(H)}=\frac{\Phi(H)}{\rv^{'}(H)}\; ,
\]
or
\be\label{O6}
\Phi(H)=\frac{(1+w)\left[\left(3H^2/8\pi\right)-\rv(H)\right]}{\left(3H/4\pi\rv^{'}(H)\right)-1}\; .
\ee
If the function $\rv(H)$ is specified, as it is always done in the papers on the $\rv(H) $ model cited in sec.~\ref{s3}, then the interaction function $\Phi(H)$ is expressed through it by this formula. If, on the other hand, one specifies the interaction $\Phi(H)$, then the formula~(\ref{O6}) turns to a first order differential equation defining $\rv(H)$. In both cases $H(t)$ is then found in quadratures from the single first order differential equation~(\ref{O5}), determining both densities as functions of time, as well as the scale factor, $a(t)\sim \exp\left[\int\limits^{t}\,H(t^{'})dt^{'}\right]$.

In several papers (see~\cite{BasSol} and the references therein) the dynamical DE density $\rv(H)$ was used in its simplest form of an even quadratic polynomial,
\be\label{Oquad}
\rv(H)=\rz+\alpha H^2,\qquad \rz,\;\alpha=\mbox{const}>0\; .
\ee
Formula~(\ref{O6}) shows that the interaction function $\Phi(H)$ is also quadratic in this case,
\[
\Phi(H)=\frac{1+w}{\left(3/8\pi\alpha\right)-1}\left\{\left[(3/8\pi)-\alpha\right]H^2-\rz\right\}\; .
\]
Remarkably, this requires $\alpha\not=3/8\pi$: if the opposite is true, then the Friedmann equation~(\ref{O2}) reduces то $\rho+\rz=0$, which can only be valid if $\rho=\rz=0$, since both densities are non-negative. The same argument shows that a physically meaningful solution requires $\alpha<3/8\pi$; otherwise at least one of the densities becomes negative.

We now extend the $\rv(H) $ model to the open and closed universe, $k=\mp1$; some particular cases of this model were considered in papers~\cite{L1}, ~\cite{L2}. Equations~(\ref{O2}), along with equation of the total energy conservation, now read:
\be\label{O7}
\frac{3}{8\pi}H^2= \rho+\rv-\frac{k}{a^2},\quad \frac{\dot a}{a}=H,\quad \dot\rho+\dot\rv= - 3H(1+w)\rho\; .
\ee
This gives three equations for the three unknown functions of time, $H,\;a$ and $\rho$, because $\rv$ is a given function of $H$, and $\dot\rv=\rv^{'}\dot H$. 

The first of the equations~(\ref{O7}) allows us to eliminate the scale factor $a(t)$ from the other two: we have
\be\label{O8}
a^2= \frac{k}{(\rho+\rv)-(3/8\pi)H^2},\qquad 
\frac{\dot a}{a}= - \frac{\dot\rho+\left[\rv^{'}-(3/4\pi)H\right]\dot H}{2\left[(\rho+\rv)-(3/8\pi)H^2\right]}\; .
\ee
So the second and third equations~(\ref{O7}) become:
\bea
\dot\rho+\left[\rv^{'}-(3/4\pi)H\right]\dot H=2H\left[(3/8\pi)H^2-(\rho+\rv)\right]\; ;\nonumber\\
\dot\rho+\rv^{'}\dot H=- 3H(1+w)\rho\; .\qquad\qquad\quad\nonumber
\eea
Solving this linear algebraic equations for $\dot\rho$ and $\dot H$, we obtain the governing system of two autonomous equations resolved with respect to the derivatives,
\bea\label{O9}
\dot H=-\frac{8\pi}{3}\left\{\left[\frac{3}{8\pi}H^2-\rv(H)\right] +(2+3w)\rho\right\}\; ;\qquad\qquad\qquad\quad\;\\
\dot\rho+\left[3(1+w)H-\frac{8\pi}{3}(2+3w)\rv^{'}(H)\right]\rho=
\frac{8\pi}{3}\left[\frac{3}{8\pi}H^2-\rv(H)\right]\rv^{'}(H)\; ,\nonumber
\eea
for the two unknown functions $H(t)$ and $\rho(t)$.

Note that the parameter $k$ designating open or closed universe case dropped out of this system. However, the first of the relations~(\ref{O8}) requires
\be\label{O10}
{k}\left[(\rho+\rv)-(3/8\pi)H^2\right]>0\; ,
\ee
which condition, as well as the usual $\rho>0$, significantly limits the set of physical solutions. 

Moreover, since, by the first equation~(\ref{O9}),
\[
\dot\rv=\rv^{'}\dot H=-\rv^{'}\frac{8\pi}{3}\left\{\left[(3/8\pi)H^2-\rv(H)\right] +(2+3w)\rho\right\}\; ,
\]
we find
\be\label{O11}
\frac{d\rv}{dV}=\frac{8\pi\rv^{'}}{9H}\left\{\left[(3/8\pi)H^2-\rv(H)\right] +(2+3w)\rho\right\}=F(\rv,\rho)\; ,
\ee
because $H=H(\rv)$. Therefore the $\rv(H)$ model for the open and closed universe ($k=\mp1$) is also a particular case of our DE--matter interaction model~(\ref{key21ph}).

If $H(t)$ and $\rho(t)$ are found from the system~(\ref{O9}), then the DE density is given by $\rv=\rv(H(t))$, and the scale factor is determined by the first equation~(\ref{O8}). In practice, a natural way to solve the system~(\ref{O9}) is to express $\rho$ form its first equation and introduce to the second one. This gives a second order autonomous differential equation for $H(t)$, which reduces, by means of a standard transformation, to a first order equation for $\dot H$ as a function of $H$. When the latter can be analytically integrated, an exact solution of the whole problem can be obtained.

\section{Two Classes of Non-Linear Interaction Laws
 Allowing for General Explicit Solutions (the Case of  a Single Matter Phase)\label{C}}
 
Here we study two interaction laws depending on both densities $\rho$ and $\rv$, for which the equations governing cosmological evolution are explicitly integrable.

First, we deal with the interaction function which is conveniently written as
\be\label{Fpropro}
 F(\rv,\rho)=\rho/f^{'}(\rv)\; ,
\ee
$f(\rv)$ being  an arbitrary function. Accordingly, the governing system~(\ref{govsaut1ph}) takes the form
\be\label{syst1}
\frac{d\rho}{d\lambda}=  -\rho\left[(1+w)+ \frac{1}{f^{'}(\rv)}\right];\quad
\frac{d\rv}{d\lambda}=  \frac{\rho}{f^{'}(\rv)}\; .
\ee
Dividing the first equation by the second one gives
\[
\frac{d\rho}{d\rv}=-1-(1+w)f^{'}(\rv)\; ,
\]
which is immediately integrated to produce the matter density as an explicit function of the density of heavy vacuum:
\be\label{roofrov}
 \rho=\rho(\rv)=r-\rv-(1+w)f(\rv)\; 
\ee
($r$ is an arbitrary constant of integration). Using this in the second equation~(\ref{syst1}) we determine the dependence of $\rv$ on $\lambda$, or on $V$:
\be\label{rovofV1}
\int\limits^{\rv}\,\frac{f^{'}(x)dx}{\rv+(1+w)f(\rv)-r}=-\lambda=\ln\frac{\Vst}{V}\; .
\ee
If this transcendental equation has a solution $\rv(V)$, then $\rho=\rho(\rv(V))$ is given by the expression~(\ref{roofrov}), and we obtain thus an exact solution of the system~(\ref{syst1}). All the solutions are described by the integrals~(\ref{roofrov}) and~(\ref{rovofV1}), so the system~(\ref{syst1}) is completely integrable. 

However, even if a solution to the equation~(\ref{rovofV1}) exists, it might not lead to a proper cosmological solution, since the latter requires the two densities to be non-negative, the matter density to vanish at large times $(V\to\infty)$, and the DE density to stay finite in the same limit. Formulating some sufficient but general enough conditions on the function $f(\rv)$ that guarantee this is rather difficult, if possible at all. It is also not easy to find a particular function $f(\rv)$ that  provides a simple enough physical solution. 

We now turn to the interaction law of the form
\be\label{Fproprov}
 F(\rv,\rho)=-\frac{(1+w)\rho\rv f^{'}(\rho)}{1+\rv f^{'}(\rho)}\; ,
\ee
where $f(\rho)$ is arbitrary. This might seem too elaborate, but it allows for an exact integration of the governing system~(\ref{govsaut1ph}), which is
\be\label{syst2}
\frac{d\rho}{d\lambda}=  -\left[(1+w)\rho+F\right]=-\frac{(1+w)\rho}{1+\rv f^{'}(\rho)};\qquad
\frac{d\rv}{d\lambda}= F(\rv,\rho)\; .
\ee
Indeed, we rewrite the expression~(\ref{Fproprov}) as 
\[
F=-\rv f^{'}(\rho)\left[(1+w)\rho+F\right]=\rv f^{'}(\rho)\frac{d\rho}{d\lambda}\;,
\]
where the last equality is implied by the first equation~(\ref{syst2}). Hence
\[
F=\rv \frac{df(\rho)}{d\lambda}\;,
\]
so the second equation~(\ref{syst2}) becomes
\[
\frac{d\rv}{d\lambda}=\rv \frac{df(\rho)}{d\lambda}\; ,
\]
and immediately integrates to give the DE density as a function of the density of matter:
\be\label{rovofro}
 \rv=\rv(\rho)=r\exp[f(\rho)]\; 
\ee
($r>0$ is a constant of integration). Using this in the first equation~(\ref{syst2}) we turn it to the equation with separable variables whose integral is a transcendental equation determining $\rho=\rho(V)$:
\be\label{roofV}
\int\limits^{\rho}\,\frac{1+r f^{'}(x)\exp[f(x)]}{x}\,dx=-(1+w)\lambda=\ln\left(\frac{\Vst}{V}\right)^{1+w}\; .
\ee
As in the previous case, any solution of this combined with the expression~(\ref{rovofro}) gives a solution to the system~(\ref{govsaut1ph}). And again, by far not any such solution makes physical sense.

\section{Behavior of Non--Singular Cosmological Solutions\\
Obtained by the Semi--Inverse Method
in the Beginning and at the End of the Expansion (the Case of  a Single Matter Phase)\label{A}}

Under the conditions~(\ref{asymph}) - (\ref{nu0}), let us check the behavior of the solution $\rv(V)$ to the resolving equation~(\ref{rvac}), derived in section~\ref{s6.3}, in the limits  $\rv\to\rz-0$ and $\rv\to\rinf+0$, i.e., at the beginning and end of the expansion.

For  the first limit we use the second representation~(\ref{asymph}) to evaluate asymptotically the integral $H(\rv)$ involved  in the equation~(\ref{rvac}), which gives:
\[
H(\rv)=-\frac{(\rho_0-\rv)^{1-\nu_0}}{h_0(1-\nu_0)},\; \nu_0\not=1; \quad
H(\rv)=-\frac{1}{h_0}\,\ln(\rho_0-\rv),\;\nu_0=1\; .\nonumber
\]
Therefore we obtain the following asymptotic forms of this equation for $\rv(V)$:
\bea\label{asrvac0}
(\rv-\rho_0)^{\nu_0}\exp\left[-\frac{(\rho_0-\rv)^{1-\nu_0}}{h_0(1-\nu_0)}\right]=(\rho_0)^{\nu_0}\left(\frac{\Vst}{V}\right)^{1+w},\quad \nu_0>1\; ;\nonumber\\
(\rho_0-\rv)^{1-1/h_0}=(\rho_0)^{1-1/h_0}\left(\frac{\Vst}{V}\right)^{1+w},\quad \nu_0=1\; .
\eea
The constants $(\rz)^{\nu_0},\;(\rz)^{1-1/h_\infty}$ are introduced here for the consistency of writing; effectively, only one arbitrary constant is present in each line. 

The left hand sides of the above equations tend to infinity when $\rv\to\rho_0-0$ (recall that in the second line $0<h_0<1$ by the condition~(\ref{nu0})). Their right hand sides match this infinity only when $V\to+0$; so the expansion necessarily starts with the zero value of the scale factor. As shown in section~\ref{s6.2}, this is only possible for the solution describing the open universe. Hence only the open non-singular cosmologies are found by the semi-inverse method under the conditions~(\ref{asymph}) - (\ref{nu0}).

Next, using the first representation~(\ref{asymph}), we calculate the asymptotics of the integral $H(\rv)$ in the limit  $\rv(V)\to\rinf+0$:
\bea
H(\rv)=\int\limits^{\rv}\,\frac{dv}{h(v)}=\frac{1}{h_\infty(1-\nu_\infty)}(\rv-\rho_\infty)^{1-\nu_\infty},\quad \nu_\infty\not=1\; ;\nonumber\\
H(\rv)=\frac{1}{h_\infty}\,\ln(\rv-\rho_\infty),\qquad\qquad\quad\;\;\; \nu_\infty=1\; .\nonumber
\eea
Equation~(\ref{rvac}) in this limit becomes thus
\bea\label{asrvacinf}
(\rv-\rho_\infty)^{\nu_\infty}\exp\left[\frac{(\rv-\rho_\infty)^{1-\nu_\infty}}{h_\infty(1-\nu_\infty)}\right]=(\rho_\infty)^{\nu_\infty}\left(\frac{\Vst}{V}\right)^{1+w},\quad \nu_\infty\not=1\; ;\nonumber\\
(\rv-\rho_\infty)^{1+1/h_\infty}=(\rho_\infty)^{1+1/h_\infty}\left(\frac{\Vst}{V}\right)^{1+w},\quad \nu_\infty=1\; .
\eea
As in the previous case, the left hand sides of the equations~(\ref{asrvacinf}) tend to infinity when $\rv(V)\to\rinf+0$ for any $\nu_\infty>0$. The right hand sides become infinite only when $V\to+\infty$, which is the right limit for $t\to\infty$. 

Equations~(\ref{asrvac0}) and~(\ref{asrvacinf}) imply the following asymptotic behavior of both densities (main terms only; $\Vst>0$ is an arbitrary constant):
\bea\label{as0}
\underline{V\to+0\;\;(t\to+0),\;\;\mbox{open universe}}\qquad\qquad\qquad\quad\qquad\qquad\qquad\nonumber\\
a)\;\mbox{for}\quad \nu_0>1,\qquad\qquad\quad\qquad\qquad\quad\qquad\qquad\quad\qquad\qquad\quad\qquad\qquad\quad\qquad\qquad\quad\qquad\nonumber\\
\rv=\rho_0-\left[(1+w)h_0(\nu_0-1)\ln(V/\Vst)\right]^{-\frac{1}{\nu_0-1}}\;,\qquad\qquad\quad\qquad\quad\nonumber\\
\rho=h_0\left[(1+w)h_0(\nu_0-1)\ln\left(V/\Vst\right)\right]^{-\frac{\nu_0}{\nu_0-1}}\;;\qquad\qquad\qquad\;\;\qquad\nonumber\\
b)\;\mbox{for}\quad \nu_0=1,\;\;0<h_0<1,
\qquad\qquad\quad\qquad\qquad\quad\qquad\qquad\quad\qquad\qquad\quad\qquad\qquad\qquad\\
\rho_0-\rv\sim\rho\sim\rho_0(V/\Vst)^{\frac{(1+w)h_0}{1-h_0}}\; .\qquad\qquad\quad\qquad\qquad\quad\qquad\qquad\nonumber
\eea
\bea\label{asinf}
\underline{V\to+\infty\;\;(t\to+\infty),\;\;\mbox{open universe}}\qquad\qquad\quad\qquad\qquad\qquad\qquad\nonumber\\
a)\;\mbox{for}\quad \nu_\infty>1,\qquad\qquad\quad\qquad\qquad\quad\qquad\qquad\quad\qquad\qquad\quad\qquad\qquad\quad\qquad\qquad\quad\qquad\nonumber\\
\rv=\rho_\infty+\left[(1+w)h_\infty(\nu_\infty-1)\ln\frac{V}{\Vst}\right]^{-\frac{1}{\nu_\infty-1}}\;,\qquad\qquad\quad\qquad\quad\nonumber\\
\rho=h_\infty\left[(1+w)h_\infty(\nu_\infty-1)\ln\frac{V}{\Vst}\right]^{-\frac{\nu_\infty}{\nu_\infty-1}}\;;\qquad\qquad\qquad\quad\qquad\quad\nonumber\\
b)\;\mbox{for}\quad \nu_\infty=1,\qquad\qquad\;\,
\qquad\qquad\quad\qquad\qquad\quad\qquad\qquad\quad\qquad\qquad\quad\qquad\qquad\qquad\quad\;\;\\
\rv-\rho_\infty\sim\rho\sim\rho_\infty (V/\Vst)^{-\frac{(1+w)h_\infty}{1+h_\infty}}\;;\quad\qquad\quad\qquad\qquad\quad\qquad\quad\nonumber\\
c)\;\mbox{for}\quad 0<\nu_\infty<1,\qquad\quad\qquad\qquad\quad\qquad\qquad\quad\qquad\qquad\quad\qquad\qquad\quad\qquad\qquad\quad\qquad\nonumber\\
\rv=\rho_\infty\left[1+(V/\Vst)^{-\frac{1+w}{\nu_\infty}}\right]\,\qquad\qquad\quad
\rho=\rho_\infty (V/\Vst)^{-(1+w)}\; .\qquad\quad\qquad\quad\nonumber
\eea
Recall that $\nu_{0,\infty}$ and $h_{0,\infty}$ are positive constants defined by formulas~(\ref{asymph}) subject to the conditions~(\ref{nu0}). Asymptotic formulas~(\ref{as0}),~(\ref{asinf}) can be converted into time dependencies using scale factor expressions~(\ref{hettlarg}) and~(\ref{hettsmall}); they are given by the equalities~(\ref{ast0}) and~(\ref{astinf}).


\end{document}